\documentclass[12pt]{article}
\usepackage{epsfig}
\begin{document}
%%%%%%%%%%%%%%%%%%%%%%%%%%%%%%%%%%%%%%%%%%%%%%%%%%%%%%%%%%%%%%%%%%%%%%%%%
\title{An alternative way to decipher the nature of the doubly charmed tetraquark $T_{cc}(3875)^+$: its antiparticle
$T_{{\bar c}{\bar c}}(3875)^-$ photoproduction off nuclei near threshold}
\author{E. Ya. Paryev\\
{\it Institute for Nuclear Research of the Russian Academy of Sciences}\\
{\it Moscow, Russia}}
%==============================================================
%%=============================================================

\renewcommand{\today}{}
\maketitle

\begin{abstract}
We study the inclusive photoproduction of $T_{{\bar c}{\bar c}}(3875)^-$ mesons (which are the antiparticles of the doubly charmed tetraquarks $T_{cc}(3875)^+$ discovered recently by the LHCb Collaboration) from nuclei in the near-threshold energy region within the nuclear spectral function approach by considering incoherent direct (${\gamma}p(n) \to D^+(D^0){T_{{\bar c}{\bar c}}(3875)^-}\Lambda^+_c$) photon--nucleon $T_{{\bar c}{\bar c}}(3875)^-$ creation processes as well as five possible different scenarios for their internal structure with the main goal of clarifying the possibility to decipher this structure (and, hence, that of $T_{cc}(3875)^+$) in photoproduction via integral and differential observables. We calculate the absolute and relative excitation functions for $T_{{\bar c}{\bar c}}(3875)^-$ production off $^{12}$C and $^{184}$W target nuclei at near-threshold photon beam energies of 30--38 GeV, the absolute differential cross sections for their production off these target nuclei at laboratory polar angles of 0$^{\circ}$--10$^{\circ}$ as well as the A and momentum dependences of the relative (transparency ratios) cross sections for $T_{{\bar c}{\bar c}}(3875)^-$ production at photon energy of 35 GeV within the adopted scenarios for the $T_{{\bar c}{\bar c}}(3875)^-$ meson intrinsic structure. We demonstrate that the absolute and relative observables considered show a certain sensitivity to these scenarios. Therefore, the measurement of such observables in future experiments at the proposed high-luminosity electron-ion colliders EIC and EicC in the US and China in the near-threshold energy region might shed light on the $T_{cc}(3875)^+$ internal structure.
\end{abstract}

\newpage

\section*{1. Introduction}

\hspace{1.5cm} The study of exotic heavy hadrons, which exhibit properties incompatible with the
predictions of the conventional quark model for quark-antiquark mesons and three-quark baryons and which
are composed of four or five quarks (and antiquarks) and are named as tetraquark and pentaquark
states, respectively, has received considerable interest in recent years and is one of the most exciting
topics of the nuclear and hadronic physics nowadays owing to the expectation to better understand perturbative
and nonperturbative aspects of QCD and to extend our knowledge about the world of "elementary" particles.
(for a recent overview of the multiquark states see Refs. [1--5]). This interest was triggered by
the discovery of $X(3872)$ resonance (also known as $\chi_{c1}(3872)$) by the Belle Collaboration in
2003 [6] as a narrow peak in the vicinity of the $D^0{\bar D}^{*0}$ (${\bar D}^0{D}^{*0}$) mass threshold in the ${J/\psi}{\pi^+}\pi^-$ invariant mass distribution in exclusive $B^{\pm} \to K^{\pm}X(3872) \to K^{\pm}({J/\psi}{\pi^+}\pi^-)$ decays.

Recently (in 2021), the LHCb Collaboration announced the first observation of the isoscalar doubly charmed tetraquark state $T_{cc}(3875)^+$ in the $D^0D^0\pi^+$ mass spectrum in the prompt production of the proton-proton
collisions at LHC [7, 8]. This state has a quark content $cc{\bar u}{\bar d}$, a mass of approximately 3875 MeV and
manifests itself as a very narrow peak in this spectrum just below the $D^0D^{*+}$ mass threshold
\footnote{$^)$Thus, its mass with respect to this threshold ($m_{D^0}+m_{D^{*+}}=3875.1$ MeV) and width are
${\delta}m_{\rm exp}=-360\pm40^{+4}_{-0}$ keV, $\Gamma=48\pm2^{+0}_{-14}$ keV when the $T_{cc}(3875)^+$ is fitted
with a unitarized Breit-Wigner shape [8].}$^)$
.
Its spin-parity quantum numbers were estimated as $J^P=1^+$ by the experiment. The $T_{cc}(3875)^+$ is the second
doubly charmed hadron that has been experimentally observed up to now. In 2017, 2018, the LHCb Collaboration reported the observation of the doubly charmed baryon $\Xi^{++}_{cc}$ with quark content $ccu$ through the non-leptonic decay modes
$\Xi^{++}_{cc} \to \Lambda^+_cK^-\pi^+\pi^+$ [9] and $\Xi^+_c\pi^+$ [10].
After the experimental discovery of the $T_{cc}(3875)^+$ meson, its internal structure has been intensively debated.
Since the mass of $T_{cc}(3875)^+$ lies just below the nominal $D^0D^{*+}$ and $D^+D^{*0}$ mass thresholds
\footnote{$^)$The latter one lies at a slightly higher energy of $m_{D^+}+m_{D^{*0}}=3876.51$ MeV than the former and difference between them is only 1.41 MeV.}$^)$, it is natural to treat it as an
extremely narrow $D^+D^{*0}$-$D ^0{D}^{*+}$ weakly bound molecular state in a relative S-wave with a very small width and binding energy, dynamically generated from the interaction of the coupled channels $D ^0{D}^{*+}$ and $D^+D^{*0}$ in the isospin $I=0$ sector (see, for example, Refs. [11--14] and those cited below). On the other hand, the $T_{cc}(3875)^+$ has possibilities to be either a compact multiquark state where the quarks and antiquarks form compact configurations with the size of a typical hadron [15--17] or a mixture of a molecular and compact tetraquark states [18, 19]
\footnote{$^)$It is worth noting that the large value of the ratio of the partial radiative decay widths of the
$\chi_{c1}(3872)$ into $\psi(2S){\gamma}$ and $J/\psi{\gamma}$ final states measured very recently by the
LHCb Collaboration makes the interpretation of the $\chi_{c1}(3872)$ state as a pure
$D^0{\bar D}^{*0}+{\bar D}^0D^{*0}$ molecule questionable and provides a strong argument in favor of a compact
component also in its structure [20].}$^)$.
All these plausible interpretations of the $T_{cc}(3875)^+$ exotic state were supported by many current theoretical studies (see references herein below). But, in spite of large efforts the full understanding of the $T_{cc}(3875)^+$ structure remains a challenging open problem.

Most of the above various interpretations are based on the analysis of the three-body hadronic and radiative decays of
$T_{cc}(3875)^+$ via intermediate states $D^0D^{*+}$ and $D^+D^{*0}$ into the channels $D^0D^0\pi^+$, $D^0D^+\pi^0$
and $D^0D^+\gamma$, using the experimental values of its mass and total decay width as inputs [21].
However, the internal structure of $T_{cc}(3875)^+$ could also be studied in high-energy proton--proton and heavy--ion collisions at hadron colliders where a large number of charm quarks is produced, since its abundance in these collisions could reflect its structure [22]. In these collisions the $T_{cc}(3875)^+$ mesons can be produced as a compact tetraquarks from the four-quark coalescence in the QGP phase at the critical temperature or as a hadronic molecules from the coalescence of charm mesons $D$ and $D^*$ formed at the end of the mixed phase [22] (cf. Ref. [23] containing the similar statement for the exotic hadron $X(3872)$) and in them their such observables as the transverse momentum distributions and total production yields are expected to be completely different for both these configurations [24].
Alternatively, the structure of $T_{cc}(3875)^+$ can also be investigated through its antiparticle
$T_{{\bar c}{\bar c}}(3875)^-$ photoproduction off nuclei at near-threshold energies in the same manner as it was done for the $X(3872)$ resonance in Ref. [25].
The advantage of photoproduction compared to the hadronic collisions is that the interpretation of data from such future experiments is clearer due to insignificant strength of initial-state photon interaction and since in it a
fewer individual exclusive elementary $T_{{\bar c}{\bar c}}(3875)^-$ meson production channels need to be accounted for.

In this work, we present the detailed predictions for the absolute and relative excitation functions for $T_{{\bar c}{\bar c}}(3875)^-$ production off $^{12}$C and $^{184}$W target nuclei as well as for the A and momentum dependences of the relative (transparency ratios) cross sections for $T_{{\bar c}{\bar c}}(3875)^-$ production from ${\gamma}A$ reactions at threshold energies obtained in the framework of the first collision model within the possible different scenarios for the $T_{{\bar c}{\bar c}}(3875)^-$ (and, hence, for the $T_{cc}(3875)^+$) meson intrinsic structure. The predictions can be tested by future high-precision measurements at electron-ion colliders EIC [26] and EicC [27] with the aim of discriminating between these scenarios.

\section*{2. Framework: direct $T_{{\bar c}{\bar c}}(3875)^-$ photoproduction mechanism}

\hspace{1.5cm} Direct $T_{{\bar c}{\bar c}}(3875)^-$ photoproduction off nuclei in the near-threshold photon beam
energy region $E_{\gamma} \le 38$ GeV of interest
\footnote{$^)$Which corresponds to the center-of-mass energies $W$ of the photon-proton system $W \le 8.5$ GeV,
or to the relatively "low" excess energies $\epsilon$ above the $D^+T_{{\bar c}{\bar c}}(3875)^-\Lambda^+_c$ production threshold $W_{\rm th}=m_{D^+}+m_{T_{{\bar c}{\bar c}}(3875)^-}+m_{\Lambda^+_c}$ ($m_{D^+}$, $m_{T_{{\bar c}{\bar c}}(3875)^-}$ and $m_{\Lambda^+_c}$ are the $D^+$, $T_{{\bar c}{\bar c}}(3875)^-$ mesons and $\Lambda^+_c$ hyperon free space masses, respectively ) $0 \le \epsilon \le 0.47$ GeV and in which the doubly anticharmed tetraquark $T_{{\bar c}{\bar c}}(3875)^-$ can be observed in ${\gamma}p$ and ${\gamma}A$ reactions at the proposed electron-ion colliders EIC [26] and EicC [27] in the US and China.}$^)$
can occur in the following elementary processes with the lowest free production threshold ($\approx$ 33.9 GeV) [28, 29]:
%formula(1)
\begin{equation}
{\gamma}+p \to D^++T_{{\bar c}{\bar c}}(3875)^-+\Lambda^+_c,
\end{equation}
%formula(2)
\begin{equation}
{\gamma}+n \to D^0+T_{{\bar c}{\bar c}}(3875)^-+\Lambda^+_c.
\end{equation}
The $D^+$, $D^0$, $T_{{\bar c}{\bar c}}(3875)^-$ mesons and $\Lambda^+_c$ hyperons, produced in these processes, are sufficiently energetic.
Thus, for example, the kinematically allowed $D^+$, $T_{{\bar c}{\bar c}}(3875)^-$ mesons and final $\Lambda^+_c$
hyperon laboratory momenta in the direct process (1), proceeding on the free target proton at rest, vary within the momentum ranges of 5.749--11.125 GeV/c, 13.748--20.066 GeV/c and 7.335--13.061 GeV/c, respectively, at incident photon beam energy of $E_{\gamma}=35$ GeV.
Since the neutron and $D^0$ meson masses are approximately equal to the proton and $D^+$ meson masses, respectively,
the kinematical characteristics of final particles, produced in the reaction (2), are close to those of
final particles in the process (1). The binding of target nucleons and their Fermi motion will distort the
distributions of the outgoing high-momentum $D^+$, $D^0$, $T_{{\bar c}{\bar c}}(3875)^-$ mesons and $\Lambda^+_c$
hyperons as well as lead to a wider accessible for them momentum intervals compared to those given above
(cf. Figs. 8 and 9 given below). Since the medium effects are expected to be reduced for high momenta [30, 31],
we will ignore the medium modifications of these final particles in the case when the reactions (1), (2)
proceed on a nucleons embedded in a nuclear target
\footnote{$^)$It should be pointed out that the properties of the $D$, $T_{cc}(3875)^+$ and $T_{{\bar c}{\bar c}}(3875)^-$ mesons and $\Lambda^+_c$ hyperons in a strongly interacting nuclear environment at low energies have been theoretically investigated, respectively, in the works [32--42], [43, 44] and [45, 46].}$^)$
.

  Then, neglecting the distortion of the incident photon at energies of interest
and describing the $T_{{\bar c}{\bar c}}(3875)^-$ meson absorption by intranuclear nucleons by the absorption cross section $\sigma_{{T_{{\bar c}{\bar c}}(3875)^-}N}$, we represent the total cross section for the production of
$T_{{\bar c}{\bar c}}(3875)^-$ mesons off nuclei from the direct processes (1), (2) as follows [25]:
%formula(3)
\begin{equation}
\sigma_{{\gamma}A\to {T_{{\bar c}{\bar c}}(3875)^-}X}^{({\rm dir})}(E_{\gamma})=I_{V}[A,\sigma_{{T_{{\bar c}{\bar c}}(3875)^-}N}]
\left<\sigma_{{\gamma}p \to D^+{T_{{\bar c}{\bar c}}(3875)^-}\Lambda^+_c}(E_{\gamma})\right>_A,
\end{equation}
where
%formula(4)
\begin{equation}
I_{V}[A,\sigma]=2{\pi}\int\limits_{0}^{R}r_{\bot}dr_{\bot}
\int\limits_{-\sqrt{R^2-r_{\bot}^2}}^{\sqrt{R^2-r_{\bot}^2}}dz
\rho(\sqrt{r_{\bot}^2+z^2})
\exp{\left[-\sigma\int\limits_{z}^{\sqrt{R^2-r_{\bot}^2}}
\rho(\sqrt{r_{\bot}^2+x^2})dx\right]},
\end{equation}
%formula(5)
\begin{equation}
\rho(r)=\rho_p(r)+\rho_n(r),\,\,\,r=\sqrt{r_{\bot}^2+z^2}\,\,\, {\rm or}~r=\sqrt{r_{\bot}^2+x^2};
\end{equation}
%formula(6)
\begin{equation}
\left<\sigma_{{\gamma}p \to D^+{T_{{\bar c}{\bar c}}(3875)^-}\Lambda^+_c}(E_{\gamma})\right>_A=
\int\int
P_A({\bf p}_t,E)d{\bf p}_tdE
\sigma_{{\gamma}p \to D^+{T_{{\bar c}{\bar c}}(3875)^-}\Lambda^+_c}(\sqrt{s^*})
\end{equation}
and
%formula(7)
\begin{equation}
  s^*=(E_{\gamma}+E_t)^2-({\bf p}_{\gamma}+{\bf p}_t)^2,
\end{equation}
%formula(8)
\begin{equation}
   E_t=M_A-\sqrt{(-{\bf p}_t)^2+(M_{A}-m_{p}+E)^{2}}.
\end{equation}
Here, $\sigma_{{\gamma}p\to D^+{T_{{\bar c}{\bar c}}(3875)^-}\Lambda^+_c}(\sqrt{s^*})$ is the "in-medium"
total cross section for the production of $T_{{\bar c}{\bar c}}(3875)^-$ mesons in process (1)
\footnote{$^)$In Eq. (3) we assume that the $T_{{\bar c}{\bar c}}(3875)^-$ production cross sections
in ${\gamma}p$ and ${\gamma}n$ reactions are the same.}$^)$
at the "in-medium" ${\gamma}p$ center-of-mass energy $\sqrt{s^*}$;
$\rho_p(r)$,  $\rho_n(r)$ ($r$ is the distance from the nucleus center)
and $P_A({\bf p}_t,E)$ are normalized to the numbers of protons $Z$, neutrons
$N$ and to unity the local proton, neutron densities and the
spectral function of target nucleus with mass number $A$ ($A=Z+N$), having mass $M_A$ and radius $R$
\footnote{$^)$The information about the spectral function of target nucleus, used in the subsequent calculations,
is given in Refs. [47, 48] and the local nucleon densities, employed in them, will be defined below.}$^)$;
${\bf p}_{t}$  and $E$ are the internal momentum and binding energy of the struck target proton
just before the collision; ${\bf p}_{\gamma}$ and $E_{\gamma}$ are the momentum and energy of the incident photons;
$m_p$ is the proton bare mass.

As in Ref. [25], we suppose that the "in-medium" cross section
$\sigma_{{\gamma}p \to D^+{T_{{\bar c}{\bar c}}(3875)^-}\Lambda^+_c}({\sqrt{s^*}})$ for $T_{{\bar c}{\bar c}}(3875)^-$ production in process (1)
is equivalent to the free space cross section\\ $\sigma_{{\gamma}p \to D^+{T_{{\bar c}{\bar c}}(3875)^-}\Lambda^+_c}({\sqrt{s(E_{\gamma})}})$, in which
the free space center-of-mass energy squared $s(E_{\gamma})$ for given photon energy $E_{\gamma}$, presented by the formula
%formula(9)
\begin{equation}
s(E_{\gamma})=W^2=(E_{\gamma}+m_p)^2-{\bf p}_{\gamma}^2=m_p^2+2m_pE_{\gamma},
\end{equation}
is replaced by the in-medium expression (7).
For the free total cross section $\sigma_{{\gamma}p \to D^+{T_{{\bar c}{\bar c}}(3875)^-}\Lambda^+_c}({\sqrt{s(E_{\gamma})}})$ no data are available up to now. Therefore,
in the photon energy range $W \le $ 8.5 GeV of interest we have adopted for it the following parametrization of the results of the calculations of this cross section here (for a cutoff parameter $\alpha=1.7$) within the approach [28], in which the ${\gamma}p \to D^+{T_{{\bar c}{\bar c}}(3875)^-}\Lambda^+_c$ reaction is studied by considering the central diffractive mechanism and $T_{{\bar c}{\bar c}}(3875)^-$ as a ${\bar D}{\bar D}^*$ molecule:
%formula(10)
\begin{equation}
\sigma_{{\gamma}p \to D^+{T_{{\bar c}{\bar c}}(3875)^-}\Lambda^+_c}({\sqrt{s(E_{\gamma})}})=2.409\left(1-\frac{s_{\rm th}}{s(E_{\gamma})}\right)^{1.745}~[{\rm pb}],
\end{equation}
where
%formula(11)
\begin{equation}
  s_{\rm th}=W_{\rm th}^2=(m_{D^+}+m_{T_{{\bar c}{\bar c}}(3875)^-}+m_{\Lambda^+_c})^2.
\end{equation}
In the energy regime $W \le $ 8.5 GeV, the cross section of the ${\gamma}p \to D^+{T_{{\bar c}{\bar c}}(3875)^-}\Lambda^+_c$ reaction can reach the values of the order of 0.05 pb [28]. Although, this cross section
is quite small here, one may hope that its measurements cold be performed in the future at the electron-ion colliders in the US [26] and China [27] (cf. also [28]). The empirical formulas (10), (11) will be used in our subsequent calculations
as a guideline for a reasonable evaluation of the $T_{{\bar c}{\bar c}}(3875)^-$ yield in ${\gamma}A$ reactions.

We discuss now the local nucleon densities, employed in our calculations of the $T_{{\bar c}{\bar c}}(3875)^-$ photoproduction on the target nuclei $^{12}_{6}$C, $^{27}_{13}$Al, $^{40}_{20}$Ca, $^{63}_{29}$Cu, $^{93}_{41}$Nb, $^{112}_{50}$Sn, $^{184}_{74}$W, $^{208}_{82}$Pb and $^{238}_{92}$U considered in the present study.
For lightest nucleus $^{12}_{6}$C we use the same proton and neutron density profiles of the harmonic oscillator
model [49, 50]. For nuclei $^{27}_{13}$Al, $^{40}_{20}$Ca and $^{63}_{29}$Cu for the proton and neutron densities, $\rho_p(r)$ and $\rho_n(r)$, we have employed in our present calculations the Woods-Saxon distributions with the same radial parameters for protons and neutrons given in Ref. [25].
For medium-weight $^{93}_{41}$Nb, $^{112}_{50}$Sn and heavy $^{184}_{74}$W, $^{208}_{82}$Pb, $^{238}_{92}$U target nuclei, we adopted for protons and neutrons the two-parameter Fermi density distributions with proton and neutron density parameters also reported in Ref. [25]. As in Ref. [25], we use here the neutron density $\rho_n(r)$ in the
'skin' form.

To go further, we have to determine the $T_{{\bar c}{\bar c}}(3875)^-$--nucleon absorption cross section
$\sigma_{{T_{{\bar c}{\bar c}}(3875)^-}N}$. It is a crucial to estimate this cross section in different theoretical pictures describing  the possible underlying structure of $T_{{\bar c}{\bar c}}(3875)^-$ for its interpretation
through the $T_{{\bar c}{\bar c}}(3875)^-$ photoproduction on nuclei since there is a direct dependence of the rate of this production on the absorption cross section $\sigma_{{T_{{\bar c}{\bar c}}(3875)^-}N}$
(cf. Eq.(3)). This rate is expected to be different for different options for the inner structure of the
$T_{{\bar c}{\bar c}}(3875)^-$. That is, in our approach the rate of the $T_{{\bar c}{\bar c}}(3875)^-$ meson
photoproduction off nuclei reflects the sensitivity to this structure. This remarkable feature can be used to better
determine/constrain the inner structure of this exotic state in the future dedicated experiments. In our present study we consider three scenarios for the hypothetical $T_{{\bar c}{\bar c}}(3875)^-$ meson based on the proposed in literature three popular configurations for the discovered doubly charmed tetraquark $T_{cc}(3875)^+$:
i) a molecular state - a loosely bound, with large spatial size characterized by the r.m.s. radius of the order of 5--7 fm or more,  $S$-wave state in the ${\bar D}^0D^{*-}$($D^-{\bar D}^{*0})$ system consisting of an anticharmed ${\bar D}^0$ and $D^{*-}$ or $D^-$ and ${\bar D}^{*0}$ mesons having the quarks contents ${\bar c}u$ and ${\bar c}d$ or ${\bar c}d$ and ${\bar c}u$, respectively, [51--65]
\footnote{$^)$Loosely bound $S$-wave charm-meson molecules like $X(3872)$ and $T_{cc}(3875)^+$ have universal
properties  determined by their binding energies [66, 67]. Accordingly, within the molecular hypothesis, the r.m.s. size $r_{T}$ of the separation of the ${\bar D}^0$ and $D^{*-}$ constituents in the $T_{{\bar c}{\bar c}}(3875)^-$ molecule
can be estimated from the relation $r_T=1/\sqrt{4\mu_0\delta_{T_{{\bar c}{\bar c}}(3875)^-}}$
by knowing the binding energy $\delta_{T_{{\bar c}{\bar c}}(3875)^-}=m_{{\bar D}^0}+m_{D^{*-}}-m_{T_{{\bar c}{\bar c}}(3875)^-}$ of the $T_{{\bar c}{\bar c}}(3875)^-$ molecule [66, 67]. Here, $\mu_0$ is the ${\bar D}^0D^{*-}$ reduced mass. Assuming that this binding energy is equal to that $\delta_{T_{cc}(3875)^+}$ of the $T_{cc}(3875)^+$
state relative to the $D^0D^{*+}$ mass threshold and taking $\delta_{T_{cc}(3875)^+}=360\pm40^{+4}_{-0}$ keV as input
in the case when an unitarized Breit-Wigner shape is used to fit the $T_{cc}(3875)^+$
[8], we find that the anticharmed mesons in the $T_{{\bar c}{\bar c}}(3875)^-$ have a sufficiently large r.m.s. separation: $r_{T}=5.3\pm0.3$ fm. The typical scale for the relative momentum between the ${\bar D}^0$ and
$D^{*-}$ mesons [66, 68], $\sqrt{2\mu_0\delta_{T_{{\bar c}{\bar c}}(3875)^-}}\simeq$25--30 MeV/c, is much smaller than their average laboratory momenta $\sim$ 8.1 and 8.8 GeV/c, respectively (see above). With these, we expect that the application of the considered below quasi-free approximation to evaluate the $T_{{\bar c}{\bar c}}(3875)^-$--nucleon absorption cross section in molecular scenario is sufficiently well justified.}$^)$,
ii) compact, $\sim$ 1 fm, diquark $[ud]$ and antidiquark $[{\bar c}{\bar c}]$ tetraquark state [24, 69--72],
and iii) a hybrid state - a state in which the $T_{{\bar c}{\bar c}}(3875)^-$ wave function contains a molecular and
compact four-quark components [18, 73].

In the "pure" molecular picture of the $T_{{\bar c}{\bar c}}(3875)^-$, whose wave function takes the
following charge-conjugation form
%formula(12)
\begin{equation}
|T_{{\bar c}{\bar c}}(3875)^->_{\rm mol}=\frac{1}{\sqrt{2}}\left(|D^-{\bar D}^{*0}>-|{\bar D}^0D^{*-}>\right)
\end{equation}
compared to that
%formula(13)
\begin{equation}
|T_{cc}(3875)^+>_{\rm mol}=\frac{1}{\sqrt{2}}\left(|D^+D^{*0}>-|D^0D^{*+}>\right)
\end{equation}
for the $T_{cc}(3875)^+$ in the isoscalar channel [52, 64, 74, 75]
\footnote{$^)$It should be pointed out that in a molecular picture (13) the $T_{cc}(3875)^+$ couples equally
to the $D^+D^{*0}$ and $D^0D^{*+}$ channels. However, in some works (see, for example, Refs. [53, 64, 65]) it was predicted that the $T_{cc}(3875)^+$ flavor wave function contains about 70\% of the $D^0D^{*+}$ and 30\% of the $D^+D^{*0}$
molecular components. Since in our model the specific flavor content of the $T_{cc}(3875)^+$ (and, respectively, of the $T_{{\bar c}{\bar c}}(3875)^-$) has no influence on the determination of the $T_{{\bar c}{\bar c}}(3875)^-$--nucleon absorption cross section in a molecular scenario, we will treat the $T_{{\bar c}{\bar c}}(3875)^-$ below in this
scenario as a $D^-{\bar D}^{*0}-{\bar D}^0D^{*-}$ molecule.}$^)$
and in which the constituents (anticharmed mesons) are weakly bound and have large spatial separation in coordinate space (see above)
the $T_{{\bar c}{\bar c}}(3875)^-$--nucleon absorption cross section $\sigma_{{T_{{\bar c}{\bar c}}(3875)^-}N}^{\rm mol}$ can be naturally estimated in the quasi-free approximation [22, 76, 77]. In this approximation, the binding energy of
the anticharmed mesons is ignored.
They are taken to be on-shell and they are considered to fly together with an average laboratory momentum of the order of 8 GeV/c each. In this approximation the
$T_{{\bar c}{\bar c}}(3875)^-$ is absorbed when a nuclear nucleon (proton or neutron) interacts (elastically or inelastically) with the ${\bar D}^0$ or with the $D^{*-}$ (or with the $D^-$ or with the ${\bar D}^{*0}$).
In each of these interactions the other anticharmed meson is a spectator.
Then, assuming that the total cross sections of the free $D^{*-}N$ and ${\bar D}^{*0}N$ interactions are the
same as those for the $D^{-}N$ and ${\bar D}^{0}N$ ones [78, 79], we can evaluate the cross section for $T_{{\bar c}{\bar c}}(3875)^-$ absorption, $\sigma_{{T_{{\bar c}{\bar c}}(3875)^-}N}^{\rm mol}$, as [25, 77]:
%formula(14)
\begin{equation}
\sigma_{{T_{{\bar c}{\bar c}}(3875)^-}N}^{\rm mol}=\sigma_{{T_{{\bar c}{\bar c}}(3875)^-}p}^{\rm mol}=\sigma_{{T_{{\bar c}{\bar c}}(3875)^-}n}^{\rm mol},
\end{equation}
where the $T_{{\bar c}{\bar c}}(3875)^-p$ and $T_{{\bar c}{\bar c}}(3875)^-n$ absorption cross sections, $\sigma_{{T_{{\bar c}{\bar c}}(3875)^-}p}^{\rm mol}$ and $\sigma_{{T_{{\bar c}{\bar c}}(3875)^-}n}^{\rm mol}$,
are given by
%formula(15)
\begin{equation}
\sigma_{{T_{{\bar c}{\bar c}}(3875)^-}p(n)}^{\rm mol}\approx\sigma_{{\bar D}^0p(n)}^{\rm el}+\sigma_{{\bar D}^0p(n)}^{\rm in}+\sigma_{D^-p(n)}^{\rm el}+\sigma_{D^-p(n)}^{\rm in}+\sigma_{D^-p \to {\bar D}^0n}(\sigma_{{\bar D}^0n \to D^-p}).
\end{equation}
Here, $\sigma_{{\bar D}^0p(n)}^{\rm el(in)}$ and $\sigma_{D^-p(n)}^{\rm el(in)}$ are the elastic (inelastic) cross
sections of the free ${\bar D}^0p$(${\bar D}^0n$) and $D^-p$($D^-n$) interactions, respectively. And,
$\sigma_{D^-p \to {\bar D}^0n}$ and $\sigma_{{\bar D}^0n \to D^-p}$ are the total cross sections of the free charge-exchange reactions $D^-p \to {\bar D}^0n$ and ${\bar D}^0n \to D^-p$, correspondingly. In the calculations we employ for them the following constants, which are relevant to the momentum regime above of 1 GeV/c of our interest:
$\sigma_{{\bar D}^0p(n)}^{\rm el}$=$\sigma_{D^-p(n)}^{\rm el}=10$ mb,
$\sigma_{{\bar D}^0p(n)}^{\rm in}$=$\sigma_{D^-p(n)}^{\rm in}=0$,
$\sigma_{D^-p \to {\bar D}^0n}=\sigma_{{\bar D}^0n \to D^-p}=12$ mb [78, 79]. Adopting these values, we get that
$\sigma_{{T_{{\bar c}{\bar c}}(3875)^-}p}^{\rm mol}=\sigma_{{T_{{\bar c}{\bar c}}(3875)^-}n}^{\rm mol}=32$ mb and, in line with Eq. (14), $\sigma_{{T_{{\bar c}{\bar c}}(3875)^-}N}^{\rm mol}=32$ mb.

In the second scenario, in which the $T_{{\bar c}{\bar c}}(3875)^-$ meson is treated as a compact ${\bar c}{\bar c}ud$
tetraquark state, its breakup cross section $\sigma_{{T_{{\bar c}{\bar c}}(3875)^-}N}^{{\rm 4{q}}}$ can be approximated
by the following relation [77]:
%formula(16)
\begin{equation}
\sigma_{{T_{{\bar c}{\bar c}}(3875)^-}N}^{{\rm 4{q}}}\approx\frac{1}{3}\sigma_{{T_{{\bar c}{\bar c}}(3875)^-}N}^{\rm mol}.
\end{equation}
For $\sigma_{{T_{{\bar c}{\bar c}}(3875)^-}N}^{\rm mol}=32$ mb we obtain that $\sigma_{{T_{{\bar c}{\bar c}}(3875)^-}N}^{{\rm 4{q}}}=10.7$ mb. It is interesting to note that this value for the quantity
$\sigma_{{T_{{\bar c}{\bar c}}(3875)^-}N}^{{\rm 4{q}}}$ is similar to that of 13.3 mb, adopted in Ref. [25]
for the high-momentum $X(3872)$--nucleon absorption cross section $\sigma_{{X(3872)}N}^{{\rm 4{q}}}$ considering
the $X(3872)$ meson as a compact tetraquark with radius $r_{4q}=0.65$ fm
and approximating the latter cross section by a geometrical cross section $\sigma_{4q}^{\rm geo}={\pi}r_{4q}^2$.

In the hybrid scenario, it is assumed that the $T_{{\bar c}{\bar c}}(3875)^-$ wave function is a linear superposition of the compact ${\bar c}{\bar c}ud$ (or 4$q$) and the molecular ${\bar D}{\bar D}^*$ components [18, 73]:
%formula(17)
\begin{equation}
|T_{{\bar c}{\bar c}}(3875)^->_{\rm hyb}=\alpha|{\bar c}{\bar c}ud>+\frac{\beta}{\sqrt{2}}\left(|D^-{\bar D}^{*0}>-|{\bar D}^0D^{*-}>\right).
\end{equation}
Here, $\alpha^2$ and $\beta^2$ represent the probabilities to find a compact and non-compact hadronic configurations,
respectively, for the normalization
%formula(18)
\begin{equation}
\alpha^2+\beta^2=1.
\end{equation}
The limiting case of $\alpha^2=0$, $\beta^2=1$ corresponds to the pure molecular interpretation of the $T_{{\bar c}{\bar c}}(3875)^-$ state, while the case of $\alpha^2=1$, $\beta^2=0$ refers to its pure compact four-quark treatment.
Following Refs. [18, 43, 44, 73], we assume that the $T_{{\bar c}{\bar c}}(3875)^-$ wave function (17) contains 80\% of the non-molecular 4$q$ component and 20\% of the molecular ${\bar D}{\bar D}^{*}$ component (($\alpha^2,\beta^2)=(0.8,0.2)$), 50\% of the compact component and 50\% of the non-compact component
(($\alpha^2,\beta^2)=(0.5,0.5)$) as well as 20\% of the non-molecular component and 80\% of the molecular component
(($\alpha^2,\beta^2)=(0.2,0.8)$).
These five options for the non-molecular and molecular probabilities of the $T_{{\bar c}{\bar c}}(3875)^-$ cover the bulk of theoretical and experimental information presently available in this field and they will allow us to see the sensitivity of the $T_{{\bar c}{\bar c}}(3875)^-$ meson photoproduction cross sections from the direct processes (1), (2) on nuclei to its internal structure. In the hybrid picture (17) of the $T_{{\bar c}{\bar c}}(3875)^-$, we can represent  the $T_{{\bar c}{\bar c}}(3875)^-$--nucleon absorption cross section $\sigma_{{T_{{\bar c}{\bar c}}(3875)^-}N}^{\rm hyb}$ in the following incoherent probability-weighted sum [25]:
%formula(19)
\begin{equation}
\sigma_{{T_{{\bar c}{\bar c}}(3875)^-}N}^{\rm hyb}=\alpha^2\sigma_{{T_{{\bar c}{\bar c}}(3875)^-}N}^{\rm 4{q}}+\beta^2\sigma_{{T_{{\bar c}{\bar c}}(3875)^-}N}^{\rm mol}.
\end{equation}
In view of the above, we set $\sigma_{{T_{{\bar c}{\bar c}}(3875)^-}N}^{\rm 4{q}}=10.7$ mb and $\sigma_{{T_{{\bar c}{\bar c}}(3875)^-}N}^{\rm mol}=32$ mb.
With these values, the $T_{{\bar c}{\bar c}}(3875)^-$ absorption cross section (19) is $\sigma_{T_{{\bar c}{\bar c}}(3875)^-N}^{\rm hyb}=15.0$, 21.35 as well as 27.7 mb for the non-molecular and molecular probabilities of the $T_{{\bar c}{\bar c}}(3875)^-$ 80\% and 20\%, 50\% and 50\% as well as 20\% and 80\%, respectively. We summarize below the results obtained above for the $T_{{\bar c}{\bar c}}(3875)^-$--nucleon absorption cross section $\sigma_{{T_{{\bar c}{\bar c}}(3875)^-}N}$ in the considered schemes for the $T_{{\bar c}{\bar c}}(3875)^-$ tetraquark inner configuration:
%formula(20)
\begin{equation}
\sigma_{{T_{{\bar c}{\bar c}}(3875)^-}N}=\left\{
\begin{array}{lll}
	10.7~{\rm mb}
	&\mbox{for tetraquark (4q) state}, \\
	&\\
    15.0~{\rm mb}
	&\mbox{for hybrid state (80\%,20\%)},\\
	&\\
    21.35~{\rm mb}
	&\mbox{for hybrid state (50\%,50\%)},\\
	&\\
    27.7~{\rm mb}
	&\mbox{for hybrid state (20\%,80\%)},\\
	&\\
    32~{\rm mb}
	&\mbox{for $D^-{\bar D}^{*0}-{\bar D}^0D^{*-}$ molecule}.
\end{array}
\right.	
\end{equation}
Therefore, the $T_{{\bar c}{\bar c}}(3875)^-$ as a ${\bar D}{\bar D}^*$ molecule has the largest absorption cross section and, hence, it is expected to be more easily absorbed than its other configurations in a nuclear medium.
We will use these values for the quantity $\sigma_{{T_{{\bar c}{\bar c}}(3875)^-}N}$ in our calculations.
For better readability of the text, in discussing the results in Section 3 we will use scenario I, II, III, IV, V
for the five lines in Eq. (20) from top to down.

Neglecting the nuclear effects considered: the target nucleon binding and Fermi motion, from Eq. (3)
we obtain the following simple expression for the total cross section
$\sigma_{{\gamma}A\to {T_{{\bar c}{\bar c}}(3875)^-}X}^{({\rm dir})}(E_{\gamma})$:
%formula(21)
\begin{equation}
\sigma_{{\gamma}A\to {T_{{\bar c}{\bar c}}(3875)^-}X}^{({\rm dir})}(E_{\gamma})=I_{V}[A,\sigma_{{T_{{\bar c}{\bar c}}(3875)^-}N}]
\sigma_{{\gamma}p \to D^+{T_{{\bar c}{\bar c}}(3875)^-}\Lambda^+_c}(\sqrt{s(E_{\gamma})}),
\end{equation}
where the elementary cross section $\sigma_{{\gamma}p \to D^+{T_{{\bar c}{\bar c}}(3875)^-}\Lambda^+_c}(\sqrt{s(E_{\gamma})})$ is given above by Eq. (10).
This expression is valid only at photon energies well above threshold
\footnote{$^)$In our case, the difference between the expressions (3) and (21) is small at
incident photon energies around 37 GeV and more, and it becomes substantial at photon energies close to the
threshold energy of 33.9 GeV as our calculations have shown
(compare magenta short-dashed and cyan short-dashed-dotted curves in figures 1--4 given below).}$^)$
.
As a measure for the
$T_{{\bar c}{\bar c}}(3875)^-$ absorption cross section $\sigma_{{T_{{\bar c}{\bar c}}(3875)^-}N}$ in nuclei and, hence, for the $T_{{\bar c}{\bar c}}(3875)^-$ intrinsic configuration we will use the so-called $T_{{\bar c}{\bar c}}(3875)^-$ transparency ratio defined as [80, 81]:
%formula(22)
\begin{equation}
S_A=\frac{\sigma_{{\gamma}A \to {T_{{\bar c}{\bar c}}(3875)^-}X}(E_{\gamma})}{A~\sigma_{{\gamma}p \to D^+{T_{{\bar c}{\bar c}}(3875)^-}\Lambda^+_c}(\sqrt{s(E_{\gamma})})},
\end{equation}
{\it i.e.} the ratio of the inclusive nuclear $T_{{\bar c}{\bar c}}(3875)^-$ photoproduction cross section divided by $A$ times the same quantity on a free proton. It should be noted that this relative observable is more
favorable compared to those based on the absolute cross sections for the aim of obtaining the information
on the $T_{{\bar c}{\bar c}}(3875)^-$ nuclear absorption [25] (and, hence, on the $T_{{\bar c}{\bar c}}(3875)^-$ internal configuration).
It is expected that the direct processes (1), (2) will be dominant in the $T_{{\bar c}{\bar c}}(3875)^-$ production in ${\gamma}A$ interactions close to threshold
\footnote{$^)$At least, at incident photon energies $\le$~35 GeV, where
the elementary processes ${\gamma}N \to D{T_{{\bar c}{\bar c}}(3875)^-}\Lambda^+_c\pi$
with one pion in the final states are expected to be suppressed in $T_{{\bar c}{\bar c}}(3875)^-$ production in ${\gamma}A$ reactions compared to those of (1), (2) due to their larger production threshold energy ($\approx$~35.1 GeV) in free ${\gamma}N$ collisions. Moreover, since the main inelastic channel in ${\gamma}N$
collisions at these photon energies is the multiplicity production of pions with rather
low energies the secondary pion-induced ${T_{{\bar c}{\bar c}}(3875)^-}$ production processes are energetically suppressed at them as well.
We expect that the full inclusive nuclear $T_{{\bar c}{\bar c}}(3875)^-$ photoproduction cross section $\sigma_{{\gamma}A \to {T_{{\bar c}{\bar c}}(3875)^-}X}(E_{\gamma})$, entering into Eq. (22), is entirely exhausted by that of Eq. (3) from the direct $T_{{\bar c}{\bar c}}(3875)^-$ production processes (1), (2) not only at photon energies $E_{\gamma} \le 35$ GeV, but also at higher energies considered in the present study.}$^)$.
Then, in the case of ignoring the considered nuclear effects and according to Eq. (21), we have:
%formula(23)
\begin{equation}
S_A=\frac{\sigma_{{\gamma}A \to {T_{{\bar c}{\bar c}}(3875)^-}X}^{({\rm dir})}(E_{\gamma})}{A~\sigma_{{\gamma}p \to D^+{T_{{\bar c}{\bar c}}(3875)^-}\Lambda^+_c}(\sqrt{s(E_{\gamma})})}=
\frac{I_{V}[A,\sigma_{{T_{{\bar c}{\bar c}}(3875)^-}N}]}{A}.
\end{equation}
With accounting for these effects, the quantity $S_A$ takes on, according to Eqs. (3) and (4), a more complex form:
%formula(24)
\begin{equation}
S_A=\frac{\sigma_{{\gamma}A \to {T_{{\bar c}{\bar c}}(3875)^-}X}^{({\rm dir})}(E_{\gamma})}{A~\sigma_{{\gamma}p \to D^+{T_{{\bar c}{\bar c}}(3875)^-}\Lambda^+_c}(\sqrt{s(E_{\gamma})})}=
\frac{I_{V}[A,\sigma_{{T_{{\bar c}{\bar c}}(3875)^-}N}]}{A}\frac{
\left<\sigma_{{\gamma}p \to D^+{T_{{\bar c}{\bar c}}(3875)^-}\Lambda^+_c}(E_{\gamma})\right>_A}{\sigma_{{\gamma}p \to D^+{T_{{\bar c}{\bar c}}(3875)^-}\Lambda^+_c}(\sqrt{s(E_{\gamma})})},
\end{equation}
which will be used in our subsequent calculations.
Usually, the transparency ratio of the meson is normalized to a light nucleus like $^{12}$C [80, 82].
With this and Eq. (24), for $T_{{\bar c}{\bar c}}(3875)^-$ we have:
%formula(25)
\begin{equation}
T_A=\frac{S_A}{S_C}=\frac{12~\sigma_{{\gamma}A \to {T_{{\bar c}{\bar c}}(3875)^-}X}^{({\rm dir})}(E_{\gamma})}{A~\sigma_{{\gamma}{\rm C} \to {T_{{\bar c}{\bar c}}(3875)^-}X}^{({\rm dir})}(E_{\gamma})}=
\frac{12~I_V[A,\sigma_{{T_{{\bar c}{\bar c}}(3875)^-}N}]}{A~I_V[{\rm C},\sigma_{{T_{{\bar c}{\bar c}}(3875)^-}N}]}
\frac{\left<\sigma_{{\gamma}p \to D^+{T_{{\bar c}{\bar c}}(3875)^-}\Lambda^+_c}(E_{\gamma})\right>_A}
{\left<\sigma_{{\gamma}p \to D^+{T_{{\bar c}{\bar c}}(3875)^-}\Lambda^+_c}(E_{\gamma})\right>_{\rm C}}.
\end{equation}
In the case when direct $T_{{\bar c}{\bar c}}(3875)^-$ production processes (1), (2) proceed on a free target
nucleons being at rest the expression (25) for the quantity $T_A$ is reduced to the following simpler form:
%formula(26)
\begin{equation}
T_A \approx \frac{12~I_V[A,\sigma_{{T_{{\bar c}{\bar c}}(3875)^-}N}]}{A~I_V[{\rm C},\sigma_{{T_{{\bar c}{\bar c}}(3875)^-}N}]}.
\end{equation}
The formulas (21)--(26) will be used in calculations of the integral observables considered at photon energies of our interest.
%%%%%%%%%%%%%%%%%%%%%%%%%%%%%%%%%%%%%%%%%%%%%%%%%%%%%%%%%%%
\begin{figure}[!h]
\begin{center}
\includegraphics[width=15.0cm]{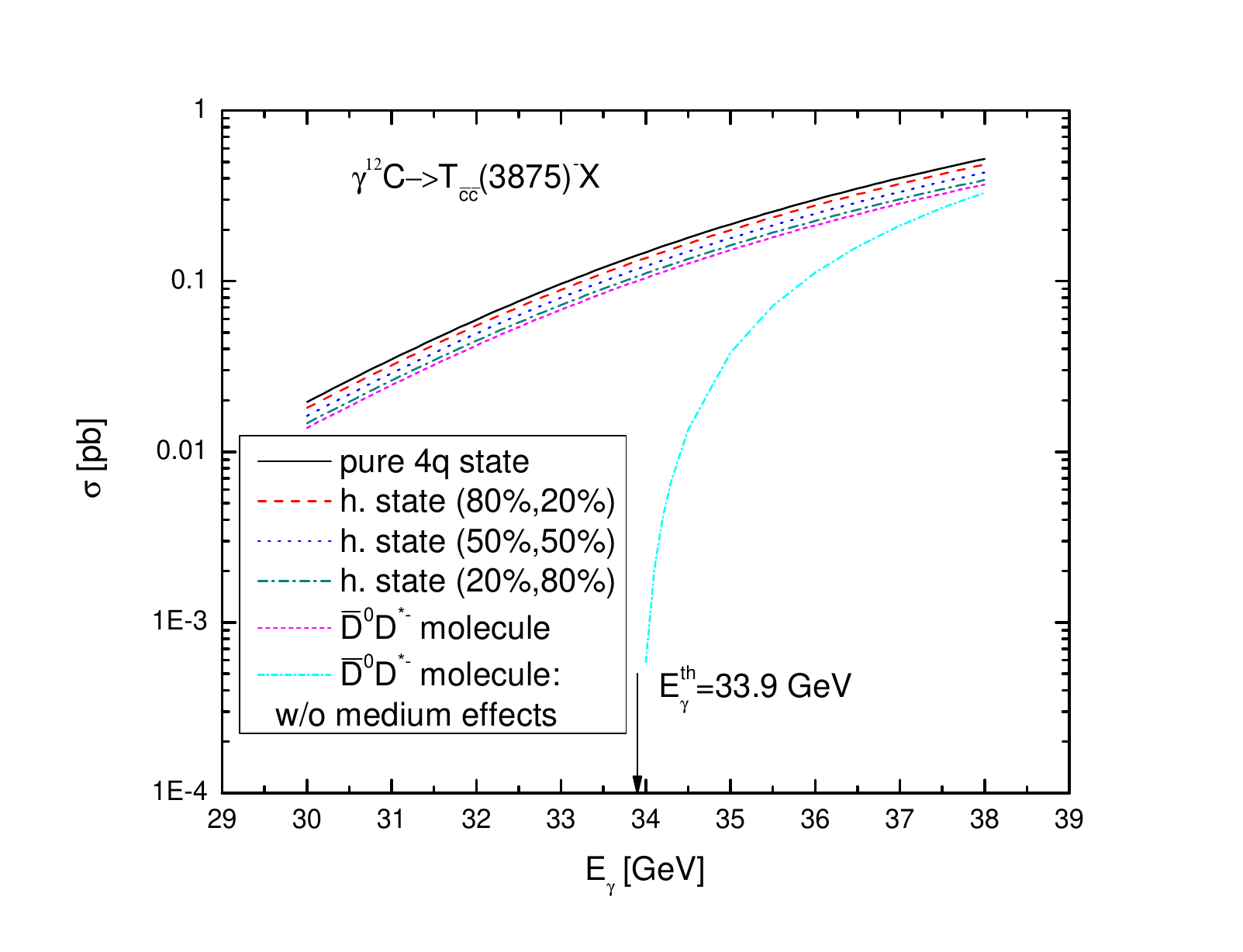}
\vspace*{-2mm} \caption{(Color online.) Excitation function for production of $T_{{\bar c}{\bar c}}(3875)^-$
mesons off $^{12}$C from primary ${\gamma}p(n) \to D^+(D^0){T_{{\bar c}{\bar c}}(3875)^-}\Lambda^+_c$ reactions
proceeding on an off-shell target nucleons and on a free ones being at rest. The curves are calculations in the scenarios, in which the $T_{{\bar c}{\bar c}}(3875)^-$ is treated as a pure tetraquark state: a compact four quark (4$q$) state, as a purely molecular state: a weakly coupled ${\bar D}^0D^{*-}$($D^-{\bar D}^{*0}$) molecule, or as a hybrid state: a mixture of the non-molecular (compact) and molecular (non-compact) components, in which there are, respectively, 80\% of the 4$q$ component and 20\% molecular component, 50\% of the 4$q$ component and 50\% molecular component as well as 20\% and 80\% of the non-molecular and molecular states. The arrow indicates the threshold energy for the $T_{{\bar c}{\bar c}}(3875)^-$ photoproduction on a free nucleon.}
\label{void}
\end{center}
\end{figure}
%%%%%%%%%%%%%%%%%%%%%%%%%%%%%%%%%%%%%%%%%%%%%%%%%%%%%%%%%%%%%%%%%%%%%%%%%%%%%%%%%%%%%%%%%%%%%
%%%%%%%%%%%%%%%%%%%%%%%%%%%%%%%%%%%%%%%%%%%%%%%%%%%%%%%%%%%%%%%%%%%%%%%%%%%%%%%%%%%%%%%%%%%%%
\begin{figure}[!h]
\begin{center}
\includegraphics[width=15.0cm]{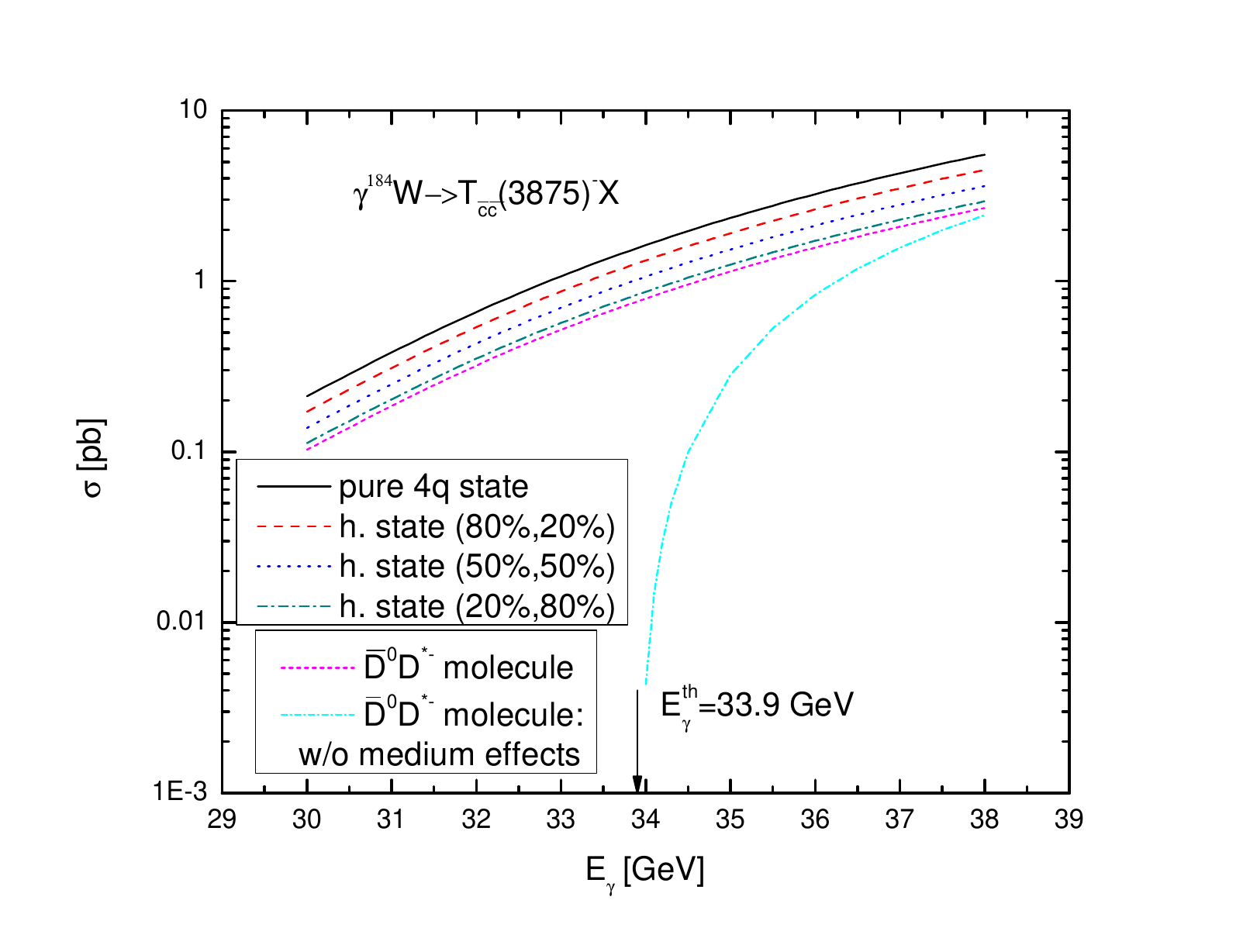}
\vspace*{-2mm} \caption{(Color online.) The same as in Fig. 1, but for the $^{184}$W target nucleus.}
\label{void}
\end{center}
\end{figure}
%%%%%%%%%%%%%%%%%%%%%%%%%%%%%%%%%%%%%%%%%%%%%%%%%%%%%%%%%%%

Before going further, we consider now the momentum-dependent inclusive differential cross section for production
of $T_{{\bar c}{\bar c}}(3875)^-$ mesons with momentum ${\bf p}_{T_{{\bar c}{\bar c}}(3875)^-}$ and total energy
$E_{T_{{\bar c}{\bar c}}(3875)^-}$ (or with momentum ${\bf p}_{T_{{\bar c}{\bar c}}^-}$ and total energy $E_{T_{{\bar c}{\bar c}}^-}$ for brevity) on nuclei in the direct processes (1) and (2) in ${\gamma}A$ reactions. According to Eqs. (3), (6) and Ref. [83], we represent this cross section as follows:
%formula(27)
\begin{equation}
\frac{d\sigma_{{\gamma}A\to {T_{{\bar c}{\bar c}}(3875)^-}X}^{({\rm dir})}(E_{\gamma},{\bf p}_{T_{{\bar c}{\bar c}}^-})}{d{\bf p}_{T_{{\bar c}{\bar c}}^-}}=
I_{V}[A,\sigma_{{T_{{\bar c}{\bar c}}(3875)^-}N}]
\left<\frac{d\sigma_{{\gamma}p \to D^+{T_{{\bar c}{\bar c}}(3875)^-}\Lambda^+_c}({\bf p}_{\gamma},{\bf p}_{T_{{\bar c}{\bar c}}^-})}{d{\bf p}_{T_{{\bar c}{\bar c}}^-}}\right>_A,
\end{equation}
where the off-shell elementary differential cross section
$d\sigma_{{\gamma}p \to D^+{T_{{\bar c}{\bar c}}(3875)^-}\Lambda^+_c}({\bf p}_{\gamma},{\bf p}_{T_{{\bar c}{\bar c}}^-})/d{\bf p}_{T_{{\bar c}{\bar c}}^-}$ for $T_{{\bar c}{\bar c}}(3875)^-$ production in process (1)
is given by (cf. Ref. [83])
%formula(28)
\begin{equation}
\frac{d\sigma_{{\gamma}p \to D^+{T_{{\bar c}{\bar c}}(3875)^-}\Lambda^+_c}({\bf p}_{\gamma},{\bf p}_{T_{{\bar c}{\bar c}}^-})}{d{\bf p}_{T_{{\bar c}{\bar c}}^-}}=\frac{\pi}{4}
\frac{\sigma_{{\gamma}p \to D^+{T_{{\bar c}{\bar c}}(3875)^-}\Lambda^+_c}(\sqrt{s^*})}
{I_3[s^*,m_{D^+},m_{T_{{\bar c}{\bar c}}(3875)^-},m_{\Lambda^+_c}]E_{T_{{\bar c}{\bar c}}^-}}
\frac{\lambda[s^*_{D^+\Lambda^+_c},m_{D^+}^{2},m_{\Lambda^+_c}^{2}]}{s^*_{D^+{\Lambda^+_c}}},
\end{equation}
assuming for simplicity that the $T_{{\bar c}{\bar c}}(3875)^-$ angular distribution in reaction (1) is
isotropic in the ${\gamma}p$ c.m.s. at photon beam energies of interest.
Here,
%FORMULA (29)
\begin{equation}
I_3[s^*,a,b,c]=\left(\frac{\pi}{2}\right)^2\int\limits_{(b+c)^2}^{(\sqrt{s^*}-a)^2}
\frac{\lambda[x,b^{2},c^{2}]}{x}\frac{\lambda[s^*,x,a^{2}]}{s^*}dx,
\end{equation}
%FORMULA (30)
\begin{equation}
\lambda(x,y,z)=\sqrt{{\left[x-({\sqrt{y}}+{\sqrt{z}})^2\right]}{\left[x-
({\sqrt{y}}-{\sqrt{z}})^2\right]}},
\end{equation}
%FORMULA (31)
\begin{equation}
s^*_{D^+{\Lambda^+_c}}=s^*+m_{T_{{\bar c}{\bar c}}(3875)^-}^2-2(E_{\gamma}+E_t)E_{T_{{\bar c}{\bar c}}^-}+2({\bf p}_{\gamma}+{\bf p}_t)
{\bf p}_{T_{{\bar c}{\bar c}}^-}
\end{equation}
and the quantity $s^*$ is defined above by Eq. (7).
At the initial photon energies of interest the $T_{{\bar c}{\bar c}}(3875)^-$ mesons are produced at very small
laboratory polar angles
\footnote{$^)$Thus, for example, the maximum angle of $T_{{\bar c}{\bar c}}(3875)^-$ meson production off a free proton at rest in reaction (1) is about 2.5$^{\circ}$ at photon energy of 35 GeV.}$^)$
. Therefore, we will calculate their momentum distribution from target nuclei considered
for the laboratory solid angle ${\Delta}{\bf \Omega}_{T_{{\bar c}{\bar c}}^-}$ = $0^{\circ} \le \theta_{T_{{\bar c}{\bar c}}^-} \le 10^{\circ}$ and $0 \le \varphi_{T_{{\bar c}{\bar c}}^-} \le 2{\pi}$.
Then, in line with Eq. (27), we can get the following expression for this distribution:
%formula(32)
\begin{equation}
\frac{d\sigma_{{\gamma}A\to {T_{{\bar c}{\bar c}}(3875)^-}X}^{({\rm dir})}
(E_{\gamma},p_{T_{{\bar c}{\bar c}}^-})}{dp_{T_{{\bar c}{\bar c}}^-}}=
\int\limits_{{\Delta}{\bf \Omega}_{T_{{\bar c}{\bar c}}^-}}^{}d{\bf \Omega}_{T_{{\bar c}{\bar c}}^-}
\frac{d\sigma_{{\gamma}A\to {T_{{\bar c}{\bar c}}(3875)^-}X}^{({\rm dir})}
(E_{\gamma},{\bf p}_{T_{{\bar c}{\bar c}}^-})}{d{\bf p}_{T_{{\bar c}{\bar c}}^-}}p_{T_{{\bar c}{\bar c}}^-}^2
\end{equation}
$$
=2{\pi}I_{V}[A,\sigma_{{T_{{\bar c}{\bar c}}(3875)^-}N}]
\int\limits_{\cos10^{\circ}}^{1}d\cos{{\theta_{T_{{\bar c}{\bar c}}^-}}}
\left<\frac{d\sigma_{{\gamma}p\to D^+{T_{{\bar c}{\bar c}}(3875)^-}{\Lambda^+_c}}(p_{\gamma},
p_{T_{{\bar c}{\bar c}}^-},\theta_{T_{{\bar c}{\bar c}}^-})}{dp_{T_{{\bar c}{\bar c}}^-}d{\bf \Omega}_{T_{{\bar c}{\bar c}}^-}}\right>_A.
$$
%%%%%%%%%%%%%%%%%%%%%%%%%%%%%%%%%%%%%%%%%%%%%%%%%%%%%%%%%%%
\begin{figure}[!h]
\begin{center}
\includegraphics[width=15.0cm]{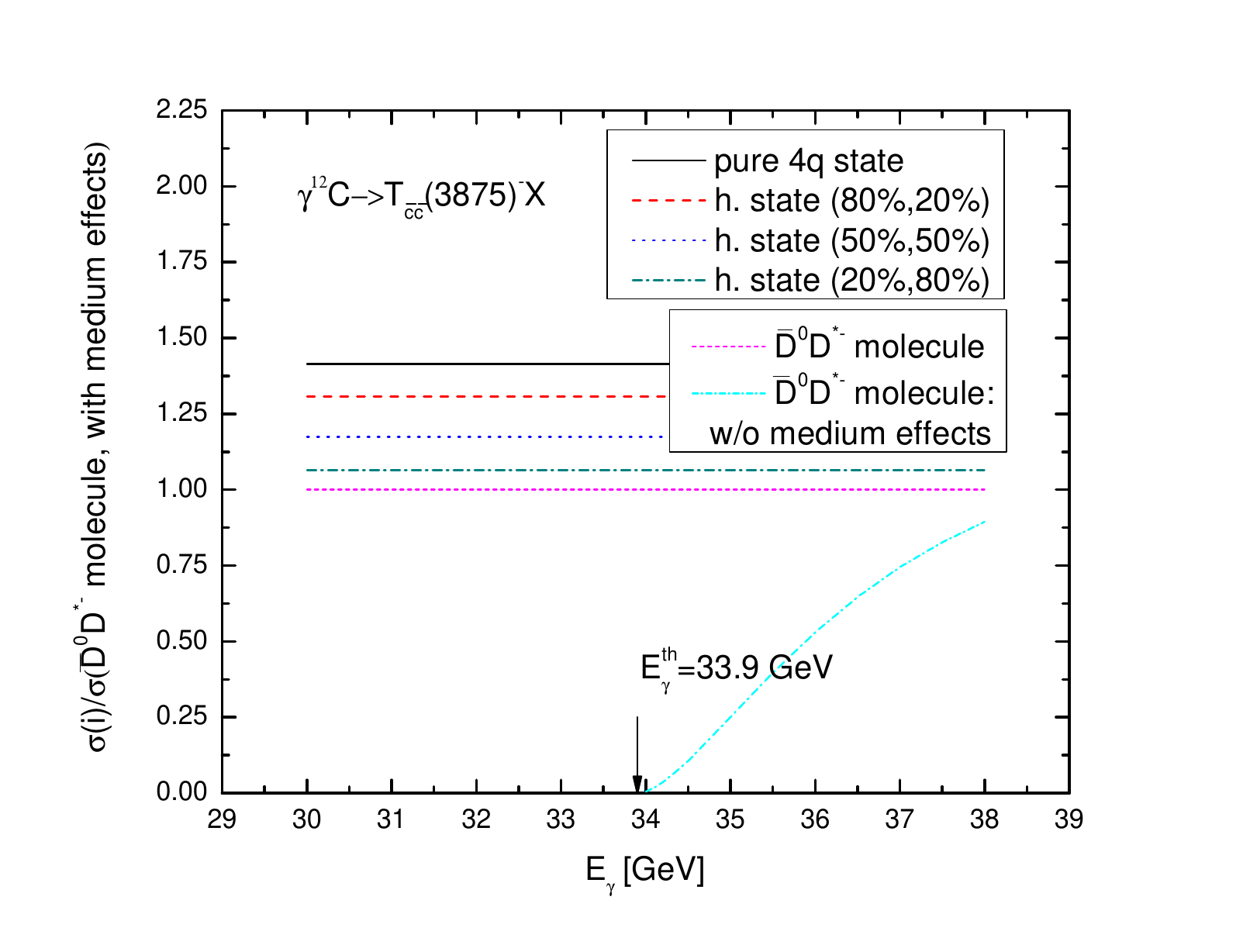}
\vspace*{-2mm} \caption{(Color online.) Ratio between the $T_{{\bar c}{\bar c}}(3875)^-$ production cross sections on $^{12}$C, shown in Fig. 1, and the cross section, calculated in the molecular scenario of $T_{{\bar c}{\bar c}}(3875)^-$ for an off-shell target nucleons, as a function of photon energy. The arrow indicates the threshold energy for the $T_{{\bar c}{\bar c}}(3875)^-$ photoproduction on a free target nucleon.}
\label{void}
\end{center}
\end{figure}
%%%%%%%%%%%%%%%%%%%%%%%%%%%%%%%%%%%%%%%%%%%%%%%%%%%%%%%%%%%%%%%%%%%%%%%%%%%%%%%%%%%%%%%%%%%%%
%%%%%%%%%%%%%%%%%%%%%%%%%%%%%%%%%%%%%%%%%%%%%%%%%%%%%%%%%%%
\begin{figure}[!h]
\begin{center}
\includegraphics[width=15.0cm]{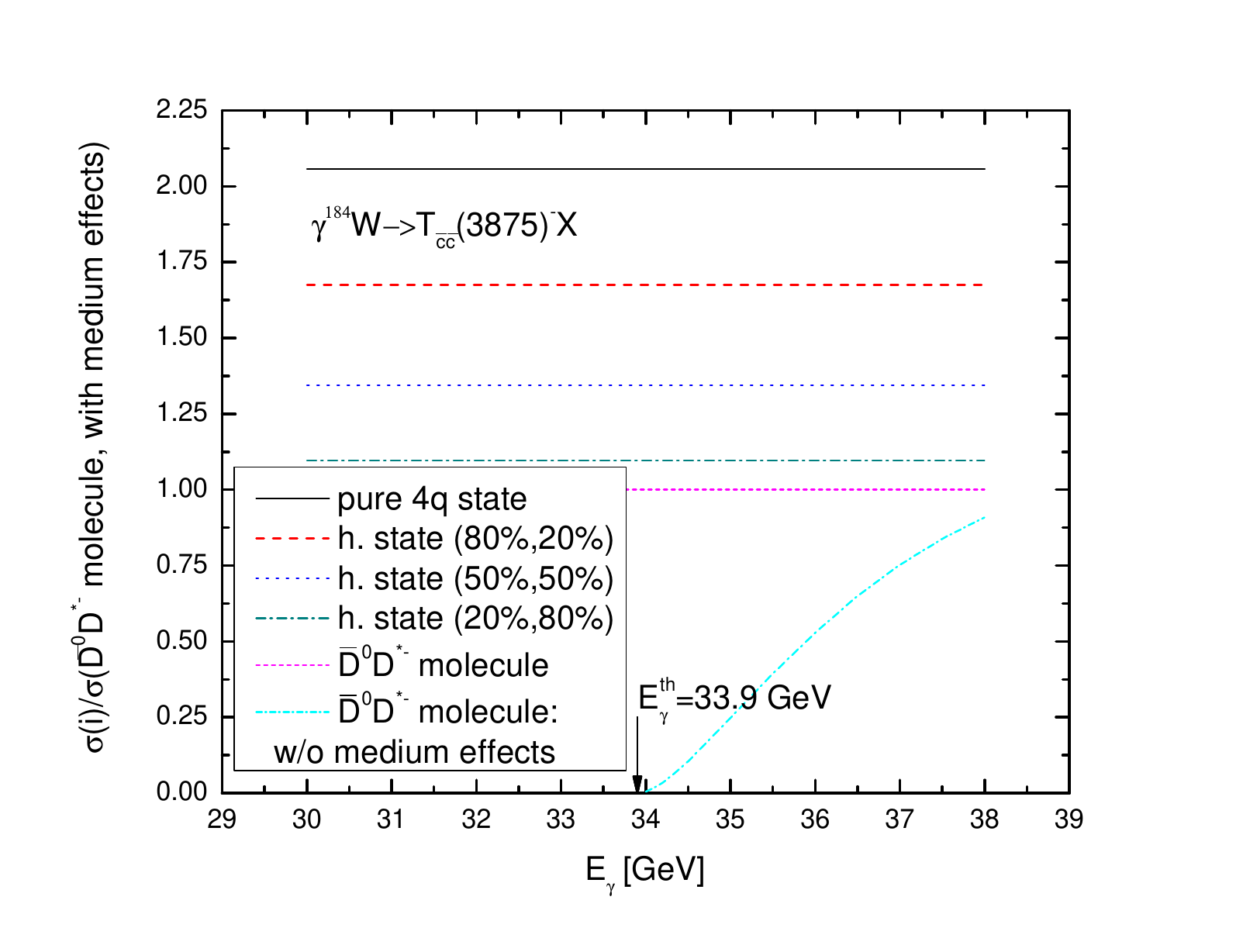}
\vspace*{-2mm} \caption{(Color online.) Ratio between the $T_{{\bar c}{\bar c}}(3875)^-$ production cross sections on $^{184}$W, shown in Fig. 2, and the cross section, calculated in the molecular scenario of $T_{{\bar c}{\bar c}}(3875)^-$ for an off-shell target nucleons, as a function of photon energy. The arrow indicates the threshold energy for the $T_{{\bar c}{\bar c}}(3875)^-$ photoproduction on a free target nucleon.}
\label{void}
\end{center}
\end{figure}
%%%%%%%%%%%%%%%%%%%%%%%%%%%%%%%%%%%%%%%%%%%%%%%%%%%%%%%%%%%%%%%%%%%%%%%%%%%%%%%%%%%%%%%%%%%%%
%%%%%%%%%%%%%%%%%%%%%%%%%%%%%%%%%%%%%%%%%%%%%%%%%%%%%%%%%%%
\begin{figure}[!h]
\begin{center}
\includegraphics[width=15.0cm]{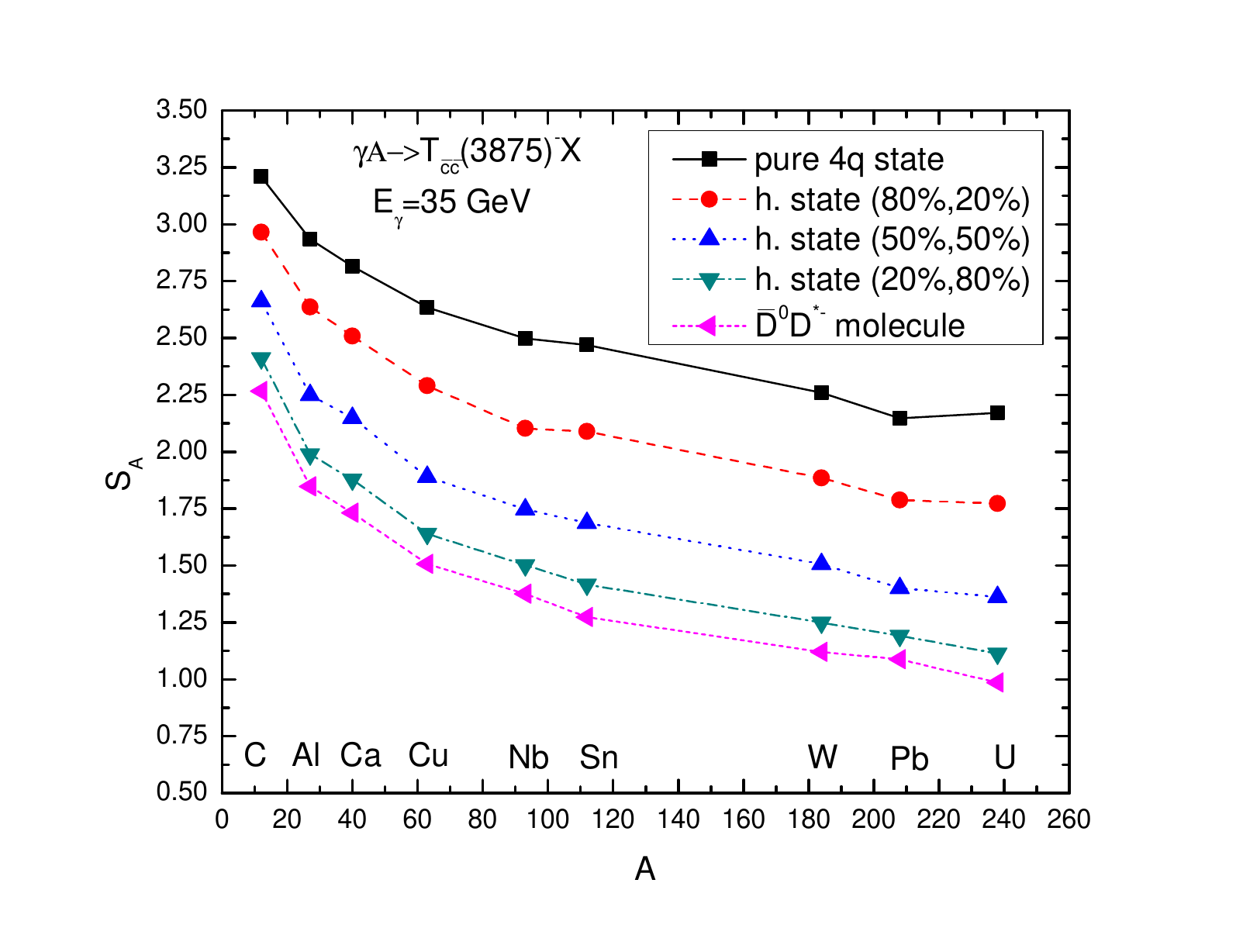}
\vspace*{-2mm} \caption{(Color online.) Transparency ratio $S_A$ for the $T_{{\bar c}{\bar c}}(3875)^-$ mesons
from primary ${\gamma}p(n) \to D^+(D^0){T_{{\bar c}{\bar c}}(3875)^-}\Lambda^+_c$ reactions proceeding on an off-shell target nucleons at initial photon energy of 35 GeV in the laboratory system as a function of the nuclear mass number $A$ in the considered theoretical pictures describing the underlying intrinsic structure of the $T_{{\bar c}{\bar c}}(3875)^-$ tetraquark state. The lines are to guide the eyes.}
\label{void}
\end{center}
\end{figure}
%%%%%%%%%%%%%%%%%%%%%%%%%%%%%%%%%%%%%%%%%%%%%%%%%%%%%%%%%%%%%%%%%%
%%%%%%%%%%%%%%%%%%%%%%%%%%%%%%%%%%%%%%%%%%%%%%%%%%%%%%%%%%%%%%%%%%
\begin{figure}[!h]
\begin{center}
\includegraphics[width=15.0cm]{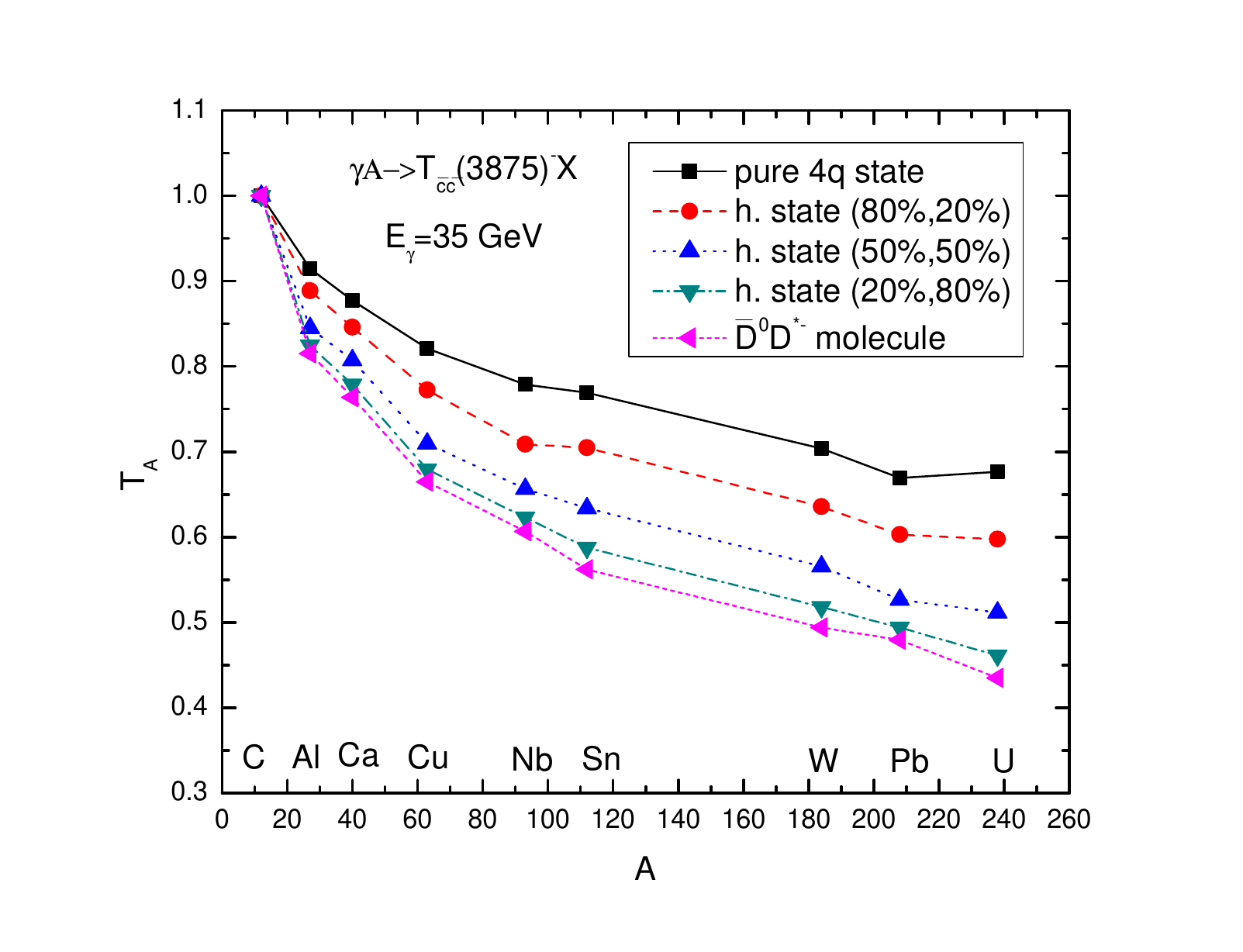}
\vspace*{-2mm} \caption{(Color online.) Transparency ratio $T_A$ for the $T_{{\bar c}{\bar c}}(3875)^-$ mesons
from primary ${\gamma}p(n) \to D^+(D^0){T_{{\bar c}{\bar c}}(3875)^-}\Lambda^+_c$ reactions proceeding on an off-shell target nucleons at initial photon energy of 35 GeV in the laboratory system
as a function of the nuclear mass number $A$ in the considered theoretical pictures describing the underlying intrinsic structure of the $T_{{\bar c}{\bar c}}(3875)^-$ tetraquark state. The lines are to guide the eyes.}
\label{void}
\end{center}
\end{figure}
%%%%%%%%%%%%%%%%%%%%%%%%%%%%%%%%%%%%%%%%%%%%%%%%%%%%%%%%%%%
%%%%%%%%%%%%%%%%%%%%%%%%%%%%%%%%%%%%%%%%%%%%%%%%%%%%%%%%%%%
\begin{figure}[!h]
\begin{center}
\includegraphics[width=15.0cm]{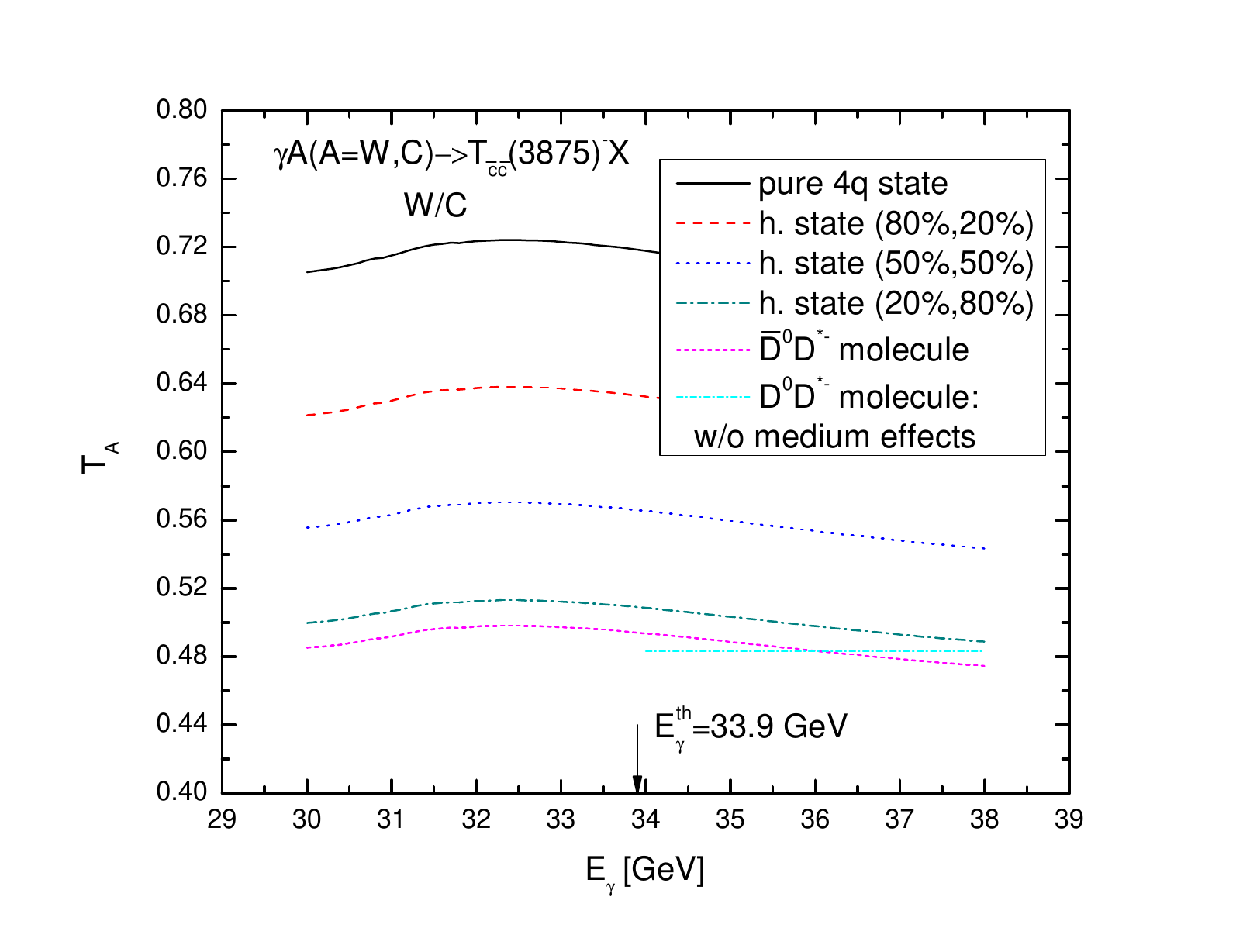}
\vspace*{-2mm} \caption{(Color online.) Transparency ratio $T_A$ for the $T_{{\bar c}{\bar c}}(3875)^-$ mesons
from primary ${\gamma}p(n) \to D^+(D^0){T_{{\bar c}{\bar c}}(3875)^-}\Lambda^+_c$ reactions proceeding on an off-shell and free target nucleons as a function of the incident photon energy for combination $^{184}$W/$^{12}$C
in the considered scenarios for the $T_{{\bar c}{\bar c}}(3875)^-$ internal structure. The arrow indicates the threshold energy for $T_{{\bar c}{\bar c}}(3875)^-$ photoproduction on a free target nucleon at rest.}
\label{void}
\end{center}
\end{figure}
%%%%%%%%%%%%%%%%%%%%%%%%%%%%%%%%%%%%%%%%%%%%%%%%%%%%%%%%%%%%%%%%%%%%%%%%%%%%%%%%%%%%%%%%%%%%%
%%%%%%%%%%%%%%%%%%%%%%%%%%%%%%%%%%%%%%%%%%%%%%%%%%%%%%%%%%%%%%%%%%%%%%%%%%%%%%%%%%%%%%%%%%%%%
\begin{figure}[!h]
\begin{center}
\includegraphics[width=15.0cm]{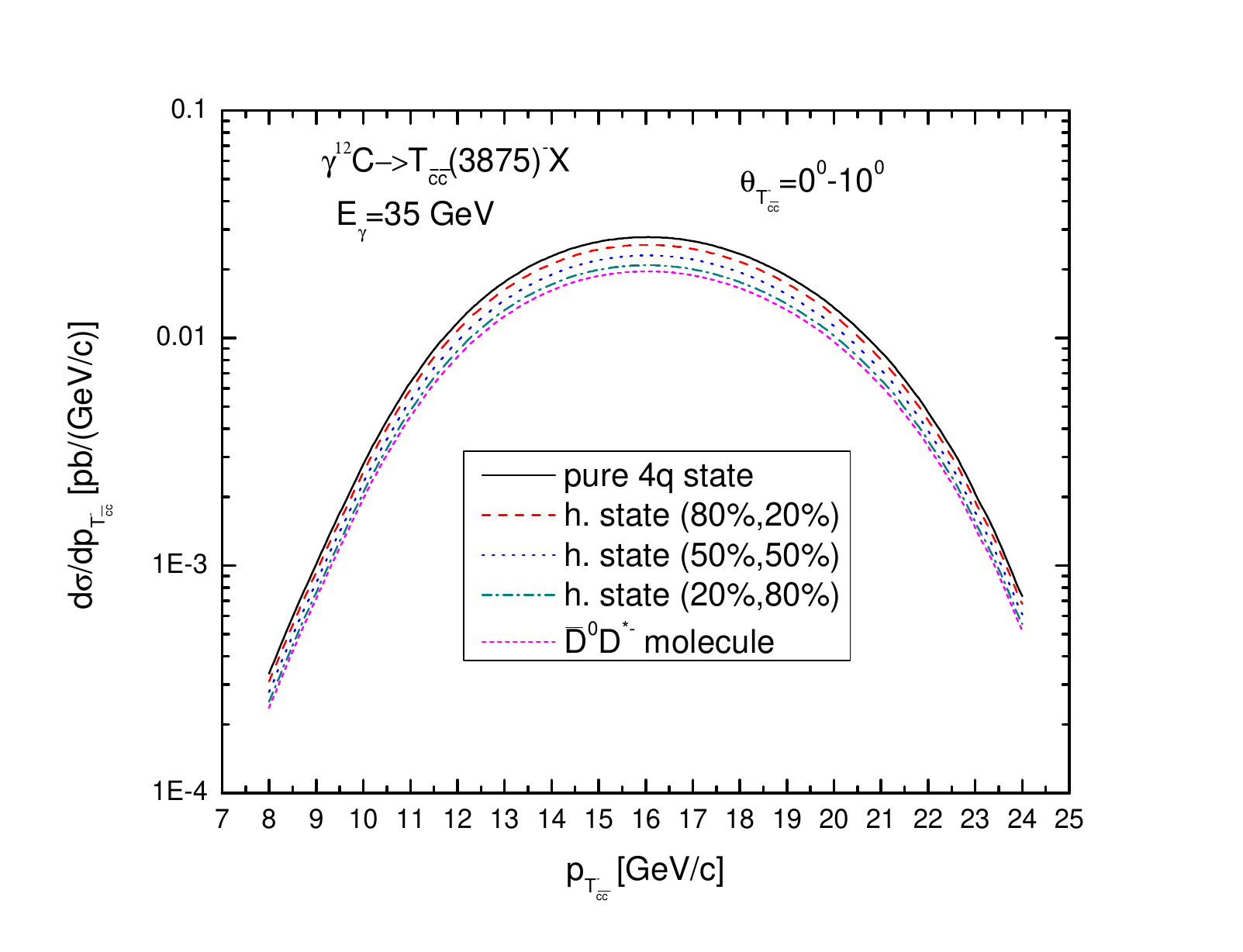}
\vspace*{-2mm} \caption{(Color online.) The direct momentum distribution of $T_{{\bar c}{\bar c}}(3875)^-$ mesons,
produced in the reaction $\gamma$$^{12}$C $\to T_{{\bar c}{\bar c}}(3875)^-X$ in the laboratory polar angular range of
0$^{\circ}$--10$^{\circ}$ and calculated in line with Eq. (32) at initial photon energy of 35 GeV in the
laboratory system. The curves are calculations in the scenarios, in which the $T_{{\bar c}{\bar c}}(3875)^-$ is treated as a purely tetraquark state: a compact four quark (4$q$) state, as a purely molecular state: a weakly coupled
${\bar D}^0D^{*-}$($D^-{\bar D}^{*0}$) molecule, or as a hybrid state: a mixture of the compact tetraquark and hadronic molecular states, in which there are, respectively, 80\% of the 4$q$ component and 20\% molecular component,
50\% of the compact component and 50\% molecular component as well as 20\% and 80\% of the non-molecular and molecular states.}
\label{void}
\end{center}
\end{figure}
%%%%%%%%%%%%%%%%%%%%%%%%%%%%%%%%%%%%%%%%%%%%%%%%%%%%%%%%%%%
%%%%%%%%%%%%%%%%%%%%%%%%%%%%%%%%%%%%%%%%%%%%%%%%%%%%%%%%%%%%%%%%%%%%%%%%%%%%%%%%%%%%%%%%%%%%%
\begin{figure}[!h]
\begin{center}
\includegraphics[width=15.0cm]{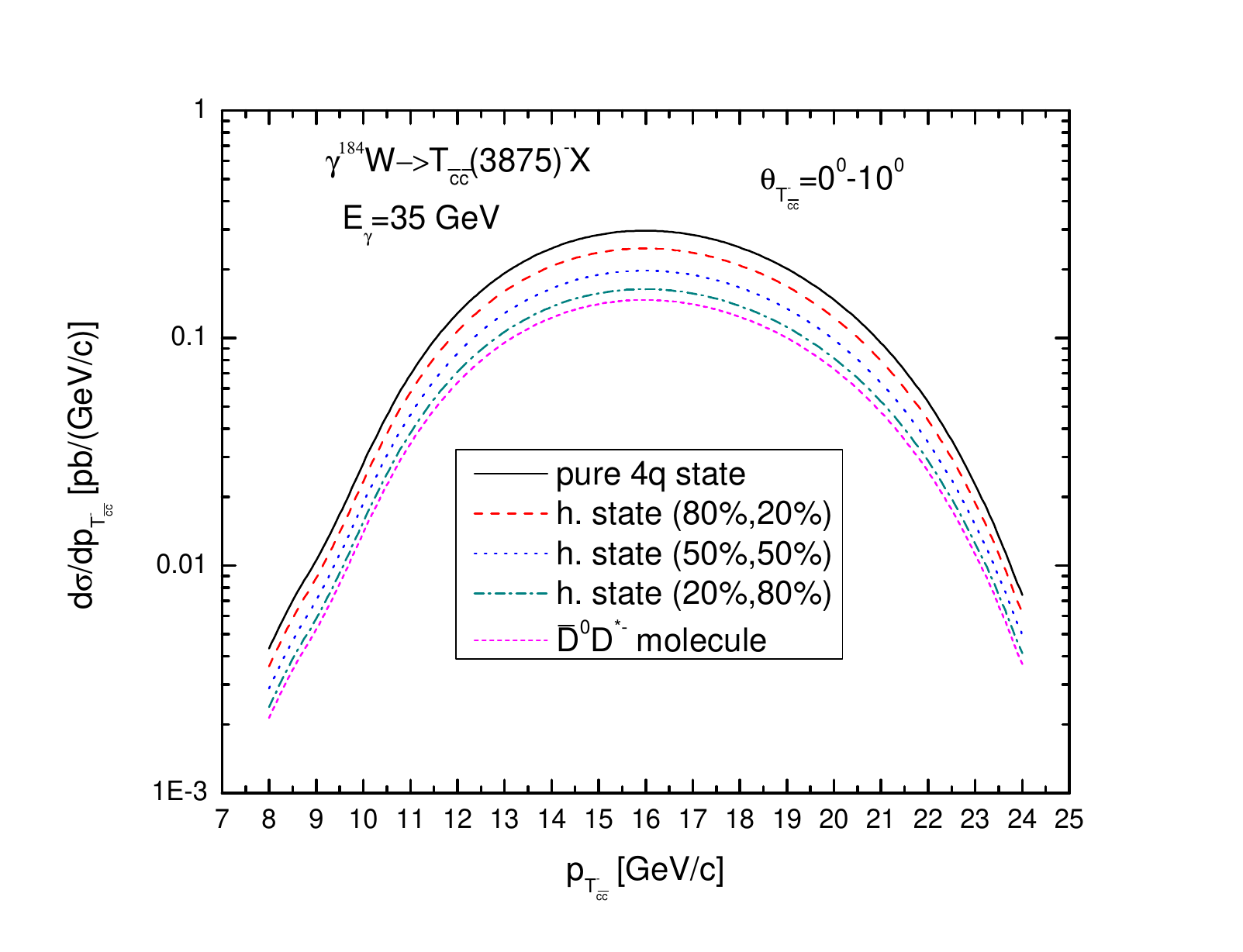}
\vspace*{-2mm} \caption{(Color online.) The same as in Fig. 8, but for the $^{184}$W target nucleus.}
\label{void}
\end{center}
\end{figure}
%%%%%%%%%%%%%%%%%%%%%%%%%%%%%%%%%%%%%%%%%%%%%%%%%%%%%%%%%%%%%%
%%%%%%%%%%%%%%%%%%%%%%%%%%%%%%%%%%%%%%%%%%%%%%%%%%%%%%%%%%%
\begin{figure}[!h]
\begin{center}
\includegraphics[width=15.0cm]{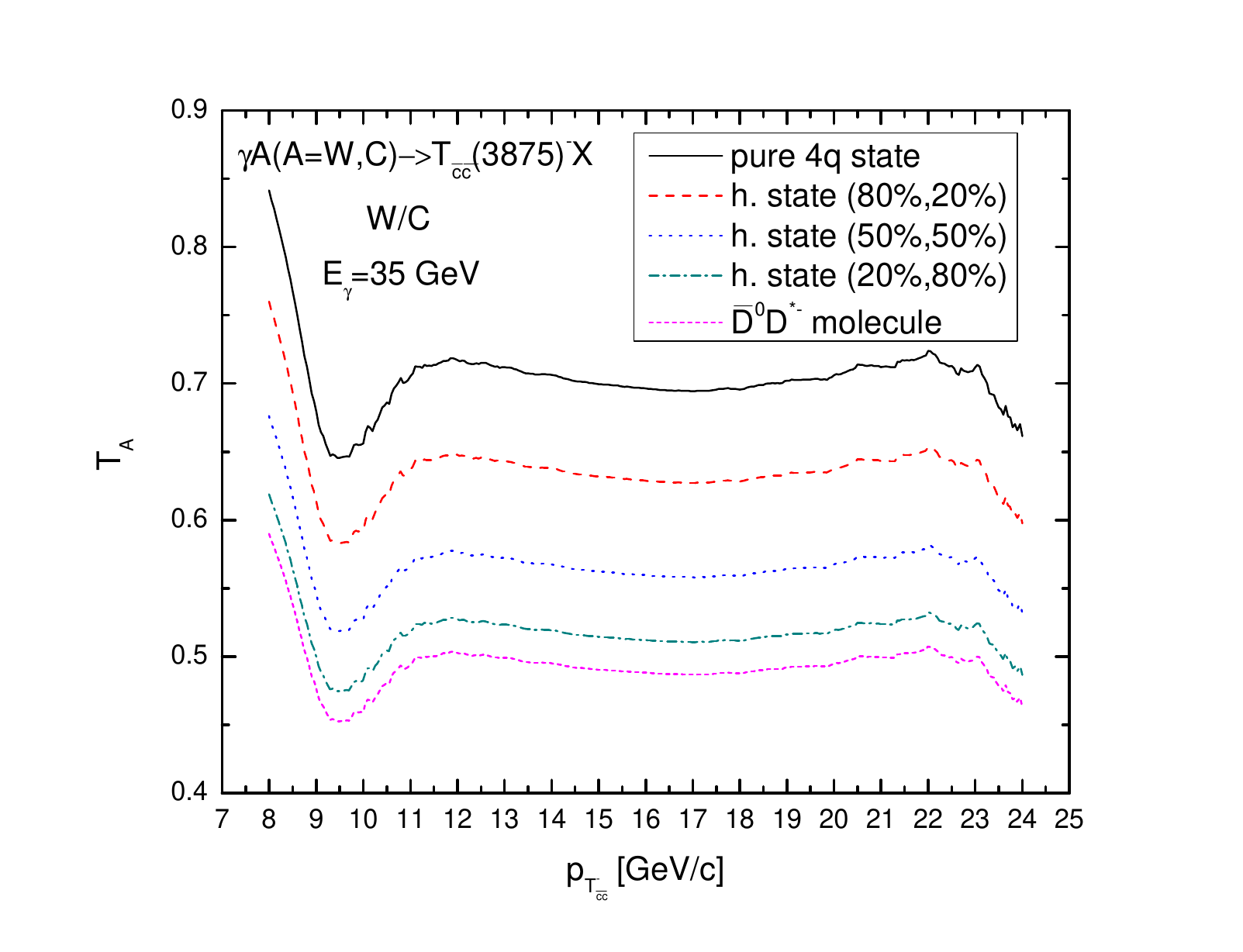}
\vspace*{-2mm} \caption{(Color online.) Transparency ratio $T_A$ for the $T_{{\bar c}{\bar c}}(3875)^-$ mesons
from primary ${\gamma}p(n) \to D^+(D^0){T_{{\bar c}{\bar c}}(3875)^-}\Lambda^+_c$ reactions proceeding on an off-shell target nucleons as a function of the $T_{{\bar c}{\bar c}}(3875)^-$ laboratory momentum for incident photon energy of 35 GeV for combination $^{184}$W/$^{12}$C in the considered scenarios for the $T_{{\bar c}{\bar c}}(3875)^-$ inner structure.}
\label{void}
\end{center}
\end{figure}
%%%%%%%%%%%%%%%%%%%%%%%%%%%%%%%%%%%%%%%%%%%%%%%%%%%%%%%%%%%%%%%%%%%%%%%%%%%%%%%%%%%%%%%%%%%%%

\section*{3. Results and discussion}

\hspace{1.5cm} We discuss now the results of our calculations in the framework of the approach outlined above.
First of all, we consider the excitation functions for production of $T_{{\bar c}{\bar c}}(3875)^-$ mesons off
$^{12}$C and $^{184}$W nuclei. They were calculated in line with Eq. (3)
for five adopted options for the $T_{{\bar c}{\bar c}}(3875)^-$ absorption cross section $\sigma_{{T_{{\bar c}{\bar c}}(3875)^-}N}$ in nuclear medium (cf. Eq. (20)) as well as in line with formula (21) for the free target nucleon
at rest with the value of $\sigma_{{T_{{\bar c}{\bar c}}(3875)^-}N}=32$ mb corresponding to the molecular picture for the $T_{{\bar c}{\bar c}}(3875)^-$, and are given in Figs. 1 and 2, respectively.
One can see that the difference between calculations with and without
accounting for the target nucleon Fermi motion (between magenta short-dashed and cyan short-dashed-dotted curves)
is sufficiently small at well above threshold photon energies $\sim$ 37--38 GeV, while at lower incident energies its impact on the $T_{{\bar c}{\bar c}}(3875)^-$ yield is substantial.
It can be seen yet that the $T_{{\bar c}{\bar c}}(3875)^-$ meson total cross sections show a definite sensitivity to the
absorption cross section $\sigma_{{T_{{\bar c}{\bar c}}(3875)^-}N}$ and, hence, to the $T_{{\bar c}{\bar c}}(3875)^-$ internal structure, for both target nuclei and for all photon energies considered.
But their absolute values are quite small even at above threshold photon energies of $\sim$ 35--38 GeV.
At these energies, they are $\sim$ 0.1--0.5 and 1--5 pb for carbon and tungsten target nuclei, correspondingly,
which is accessible at the proposed electron-ion colliders [26, 27] (see below).
For the heavy target nucleus $^{184}$W, there are experimentally distinguishable
differences between the considered scenarios for the $T_{{\bar c}{\bar c}}(3875)^-$--nucleon absorption cross section or for the $T_{{\bar c}{\bar c}}(3875)^-$ intrinsic configuration, namely: between the scenarios V and IV, IV and III, III and II, II and I.
They are $\sim$ 12, 21, 25, 20\% for all considered photon energies. For the light target nucleus $^{12}$C, the sensitivity of the cross sections to these quantities becomes somewhat lower and, respectively, the same differences as above become somewhat smaller. They are $\sim$ 6, 11, 11, 8\%. Measurement of such small effects is a difficult experimental challenge.
On the other hand, the differences between calculations corresponding to the molecular configuration and hybrid configuration with the non-molecular and molecular probabilities of 50\% and 50\%, hybrid configuration with the non-molecular and molecular probabilities of 50\% and 50\% and pure 4$q$ compact configuration for the $T_{{\bar c}{\bar c}}(3875)^-$ are yet not so small. They are $\sim$ 18, 20\%. This means that, at least, these configurations could be tested in the future $T_{{\bar c}{\bar c}}(3875)^-$ measurements on carbon and light target nuclei like carbon as well.
To motivate such measurements at the future colliders EicC and EIC, it is desirable to estimate the $T_{{\bar c}{\bar c}}(3875)^-$ production rates (the event numbers) in the ${\gamma}^{12}$C and ${\gamma}^{184}$W reactions.
For this purpose, we translate the $T_{{\bar c}{\bar c}}(3875)^-$ photoproduction total cross sections, reported above, into the expected event numbers for the $T_{{\bar c}{\bar c}}(3875)^-$ from the reactions
${\gamma}^{12}{\rm C}(^{184}{\rm W}) \to {T_{{\bar c}{\bar c}}(3875)^-}X$, $T_{{\bar c}{\bar c}}(3875)^- \to {\bar D}^0{\bar D}^0\pi^-$, ${\bar D}^0 \to K^+\pi^-$, ${\bar D^0} \to K^+\pi^-$.
To estimate the total numbers of the $T_{{\bar c}{\bar c}}(3875)^-$ events in a one-year run at EicC, one needs
to multiply the above $T_{{\bar c}{\bar c}}(3875)^-$ photoproduction total cross sections of 0.1--0.5 and 1--5 pb
on the carbon and tungsten target nuclei, respectively, by the integrated luminosity of 60 fb$^{-1}$ [27, 84]
as well as by the detection efficiency and by the appropriate branching ratios
$Br[T_{{\bar c}{\bar c}}(3875)^- \to {\bar D}^0{\bar D}^0\pi^-]\approx$ 60\% and
$Br[{\bar D}^0 \to K^+\pi^-]\approx$ 4\%
\footnote{$^)$We suppose here that $Br[T_{{\bar c}{\bar c}}(3875)^- \to {\bar D}^0{\bar D}^0\pi^-]=Br[T_{c c}(3875)^+ \to D^0D^0\pi^+]\approx$ 60\% [18, 55, 58, 85] and $Br[{\bar D}^0 \to K^+\pi^-]=Br[D^0 \to K^-\pi^+]\approx$ 4\% [86].}$^)$
.
With a realistic 50\% detection efficiency [27],
we estimate about of 3--15 and 30--150 events per year for the $T_{{\bar c}{\bar c}}(3875)^-$ signal in the cases of the $^{12}$C and $^{184}$W target nuclei, respectively. The event numbers for
$T_{{\bar c}{\bar c}}(3875)^- \to {\bar D}^0(\to K^+\pi^-){\bar D}^0(\to K^+\pi^-)\pi^-$ at EIC
are about of 15--75 and 150--750 with an integrated luminosity of 300 fb$^{-1}$ [26, 84].
We see that a sufficiently large number of $T_{{\bar c}{\bar c}}(3875)^-$ events could be observed (especially at
EIC collider) at well above threshold photon energies of 35--38 GeV.
Thus, the electron-ion colliders EicC and EIC provide a good platform to study the nature of the $T_{{\bar c}{\bar c}}(3875)^-$ (and, respectively, of the $T_{cc}(3875)^+$) in near-threshold photonuclear reactions.

To gain further insight, we show in Figs. 3 and 4 the energy dependences of the
ratios of the cross sections, presented in Figs. 1 and 2, to the cross section, calculated in the molecular picture for the $T_{{\bar c}{\bar c}}(3875)^-$ for an off-shell target nucleons, on the same linear scale for $^{12}$C and $^{184}$W target nuclei, respectively. It is clearly seen that for the heavy $^{184}$W target nucleus
the sensitivity of the $T_{{\bar c}{\bar c}}(3875)^-$ meson yield to variations in its intrinsic structure is
significantly stronger than that in the case of light $^{12}$C target nucleus and it would definitely be possible for $^{184}$W nucleus to distinguish between the results corresponding to all considered choices for this structure at all initial photon energies. At the same time, for light $^{12}$C target nucleus we see a big enough distinctions between
the curves corresponding to the pure molecular configuration, hybrid configuration with the compact and non-compact probabilities of 50\% and 50\% and pure compact tetraquark configuration for the $T_{{\bar c}{\bar c}}(3875)^-$. On the basis of the foregoing, we can conclude that the structure of $T_{{\bar c}{\bar c}}(3875)^-$ mesons could be studied at the future electron-ion colliders via the energy dependence of their absolute (and relative) total production cross sections in inclusive near-threshold photonuclear reactions. Figures 3 and 4 also show clearly that the difference between the absolute $T_{{\bar c}{\bar c}}(3875)^-$ photoproduction cross sections, calculated in the pure molecular picture for the $T_{{\bar c}{\bar c}}(3875)^-$ with allowance for the impact of the binding of intranuclear nucleons and their Fermi motion on the direct processes (1), (2) and without it (cf. Eqs. (3) and (21)) becomes indeed small only for above-threshold photon energies $\sim$ 37--38 GeV.

Additional information about the sensitivity of the cross sections to the internal structure of the
$T_{{\bar c}{\bar c}}(3875)^-$ meson is contained in Figs. 5 and 6, where the A--dependences of the transparency ratios $S_A$ and $T_A$ from the direct reaction channels (1), (2) in ${\gamma}A$
($A=$$^{12}$C, $^{27}$Al, $^{40}$Ca, $^{63}$Cu, $^{93}$Nb, $^{112}$Sn, $^{184}$W, $^{208}$Pb, and $^{238}$U)
collisions, calculated for the photon beam energy of 35 GeV in line with Eqs. (24) and (25), respectively,
and for the values of the $T_{{\bar c}{\bar c}}(3875)^-$--nucleon absorption cross section $\sigma_{T_{{\bar c}{\bar c}}(3875)^-N}$ given above by Eq. (20)
\footnote{$^)$It should be noted that the feasibility of the experimental determination of the width of a $D$ meson
in a nuclear medium by using the method of the nuclear transparency $T_A$ has been studied in recent publication [87].}$^)$
.
It is seen that the transparency ratio $S_A$ depends more strongly on this cross section
or on the adopted interpretation of $T_{{\bar c}{\bar c}}(3875)^-$ than that for the quantity $T_A$.
Thus, there are a sizeable changes $\sim$ 8, 13, 17, 11\% in the ratio $S_A$
between calculations corresponding to the cases when the loss of $T_{{\bar c}{\bar c}}(3875)^-$ mesons in nuclear
matter is determined by their absorption cross sections of 32 and 27.7 mb, 27.7 and 21.35 mb, 21.35 and 15 mb, 15 and 10.7 mb, respectively, (cf. Eq. (20)) already for comparably "light" nuclei ($^{27}$Al, $^{40}$Ca).
For the medium-mass ($^{93}$Nb, $^{112}$Sn) and heavy ($^{184}$W, $^{238}$U) target nuclei these changes are
larger and already are measurable. They are about 10, 18, 22, 18\% and 13, 21, 27, 21\%, respectively.
We see that the highest sensitivity of the quantity $S_A$ to the $T_{{\bar c}{\bar c}}(3875)^-$
structure is observed, as is expected, for heavy target nuclei.
For the quantity $T_A$ the analogous changes are smaller and are almost insignificant. They are about
2, 3, 5, 3\%, 4, 7, 10, 10\% and 6, 10, 15, 12\%, correspondingly, in the cases of "light", medium-mass
and heavy target nuclei indicated above.
Therefore, we conclude that the observation, at least, of the A dependence of the transparency ratio $S_A$, especially for large mass numbers $A$, in the future high-precision photoproduction experiments at electron-ion colliders
opens the possibility to discriminate between all considered inner configurations of the $T_{{\bar c}{\bar c}}(3875)^-$ tetraquark state.
At the same time, the precise $T_{{\bar c}{\bar c}}(3875)^-$ photoproduction data on the A dependence of the transparency ratio $T_A$ in the range of large A, obtained in such experiments, could also additionally help to distinguish, at least, between a pure compact tetraquark state and compact tetraquark-molecule mixture with the compact and molecular probabilities $\sim$ 50\% and 50\%, compact tetraquark-molecule mixture with the compact and molecular probabilities $\sim$ 50\% and 50\% and a pure molecular configuration of the $T_{{\bar c}{\bar c}}(3875)^-$ exotic state.

In addition to the A dependence of the transparency ratio $T_A$ at photon energy of 35 GeV presented in Fig. 6, we have investigated the initial photon energy dependence of this ratio for the $^{184}$W/$^{12}$C combination. This dependence is shown in Fig. 7. It was calculated in line with Eq. (25) for five adopted options for the $T_{{\bar c}{\bar c}}(3875)^-$ meson intrinsic structure and for the off-shell target nucleons as well as in line with simple formula (26) for its pure molecular interpretation and for the free target nucleons at rest. One can see that there is a certain
sensitivity of the transparency ratio $T_A$ to various interpretations of $T_{{\bar c}{\bar c}}(3875)^-$ considered
at all studied photon energies. At these energies, the sensitivity is similar to that available in Fig. 6 for heavy nuclei. Inspection of this figure tells us yet that the simple formula (26) describes well the quantity $T_A$ at above
threshold photon energies (at energies $E_{\gamma} > $ 33.9 GeV), where (and at subthreshold energies $E_{\gamma} < $ 33.9 GeV) it exhibits practically a flat behavior. This behavior of the transparency ratio $T_A$ can also be used for discriminating between possible scenarios for the $T_{{\bar c}{\bar c}}(3875)^-$ internal structure.
Thus, it is optimistic that this relative (also integral) observable can, additionally, be used to help
determine this structure.

We discuss now the $T_{{\bar c}{\bar c}}(3875)^-$ differential observables.
Figs. 8 and 9 show the absolute $T_{{\bar c}{\bar c}}(3875)^-$ meson momentum distributions from the direct
processes (1), (2), respectively, in $\gamma$$^{12}$C and $\gamma$$^{184}$W reactions, determined in line with
Eq. (32) for laboratory polar angles of 0$^{\circ}$--10$^{\circ}$, for photon beam energy of 35 GeV and
for five adopted values of the $T_{{\bar c}{\bar c}}(3875)^-$--nucleon absorption cross section
in the considered interpretations of $T_{{\bar c}{\bar c}}(3875)^-$.
Our results indicate that the differential cross sections for producing the $T_{{\bar c}{\bar c}}(3875)^-$ mesons
on $^{12}$C nucleus are roughly one order of magnitude smaller than those on $^{184}$W nucleus and, therefore,
they have a lesser chance to be observed (at EIC).
The latter ones have, according to estimates given above, a rather measurable (at EIC) strength $\sim$ 0.1--0.3 pb/(GeV/c) in the central momentum region of 13--19 GeV/c and also show a rather sizeable variations
\footnote{$^)$Which are similar to those shown in Fig. 2.}$^)$
, when going from the $T_{{\bar c}{\bar c}}(3875)^-$--nucleon absorption cross section value of 10.7 to 32.0 mb.
Such behavior of the $T_{{\bar c}{\bar c}}(3875)^-$ production differential cross sections on tungsten target nucleus
can also be used to decipher the $T_{{\bar c}{\bar c}}(3875)^-$ inner structure from comparison our model
calculations with future experimental data.

 Finally, in Fig. 10 we show the "differential" transparency ratio $T_A$ for target combination W/C for
$T_{{\bar c}{\bar c}}(3875)^-$ mesons produced in the direct reaction channels (1), (2) at laboratory angles
of 0$^{\circ}$--10$^{\circ}$ by 35 GeV photons as a function of the $T_{{\bar c}{\bar c}}(3875)^-$ laboratory
momentum. This ratio is calculated in line with Eq. (25) by using in it the differential cross sections presented
in Figs. 8 and 9 instead of the total cross sections. One can see that the sensitivity of the transparency ratio
$T_A$ to the adopted configurations of $T_{{\bar c}{\bar c}}(3875)^-$ is similar to that available in Fig. 7 for
the "integral" quantity $T_A$ considered as a function of photon energy. The transparency ratio $T_A$ exhibits dips
at momenta $\sim$ 9--10 GeV/c. This can be explained by the fact that the differential cross sections for
$T_{{\bar c}{\bar c}}(3875)^-$ production on $^{184}$W nucleus "are bent down" at these momenta (see Fig. 9)
due to the off-shell kinematics of the direct photon-nucleon collision and the role played by the nucleus-related
effects such as the target nucleon binding and Fermi motion, encoded in the nuclear spectral function
$P_A({\bf p}_t,E)$. The spectral functions for $^{12}$C and $^{184}$W, employed in the calculations,
are different [48]. This behavior of the "differential" transparency ratio $T_A$ can additionally be adopted for discriminating between possible schemes for the $T_{{\bar c}{\bar c}}(3875)^-$ internal structure.

In conclusion, the absolute total and differential cross sections for production of $T_{{\bar c}{\bar c}}(3875)^-$
mesons in near-threshold photonuclear reactions as well as their relative (transparency ratios) yields might allow one to shed light on the $T_{{\bar c}{\bar c}}(3875)^-$ (and, respectively, on the $T_{cc}(3875)^+$) intrinsic configuration.

\section*{4. Conclusions}

\hspace{1.5cm} Motivated by the experimental discovery of a doubly charmed tetraquark state $T_{cc}(3875)^+$, in this work we have studied theoretically the inclusive photoproduction of $T_{{\bar c}{\bar c}}(3875)^-$ mesons (which are the antiparticles of the $T_{cc}(3875)^+$) from nuclei in the near-threshold energy region within the nuclear spectral function approach by considering incoherent direct (${\gamma}p(n) \to D^+(D^0){T_{{\bar c}{\bar c}}(3875)^-}\Lambda^+_c$) photon--nucleon $T_{{\bar c}{\bar c}}(3875)^-$ creation processes as well as five possible different scenarios for their internal structure with the main goal of clarifying the possibility to decipher this structure (and, hence, that of $T_{cc}(3875)^+$) in photoproduction via integral and differential observables. We have calculated the absolute and relative excitation functions for $T_{{\bar c}{\bar c}}(3875)^-$ production off $^{12}$C and $^{184}$W target nuclei at near-threshold photon beam energies of 30--38 GeV, the absolute differential cross sections for their production off these target nuclei at laboratory polar angles of 0$^{\circ}$--10$^{\circ}$ as well as the A and momentum dependences of the relative (transparency ratios) cross sections for $T_{{\bar c}{\bar c}}(3875)^-$ production at photon energy of 35 GeV within the adopted scenarios for the $T_{{\bar c}{\bar c}}(3875)^-$ meson intrinsic structure. We show that the absolute and relative observables considered reveal a certain sensitivity to these scenarios. Hence, they may be an important tool to clarify it and that of the $T_{cc}(3875)^+$. The measurements of these observables could be performed in the future experiments at the proposed high-luminosity electron-ion colliders EIC and EicC in the US and China.
\\

%%%%%%%%%%%%%%%%%%%%%%%%%%%%%%%%%%%%%%%%%%%%%%%%%%%%%%%%%%%%%%%%
\end{document}